# Random sampling of colourings of sparse random graphs with a constant number of colours

Charilaos Efthymiou [*]      Paul G. Spirakis [†]

**Abstract**

In this work we present a simple and efficient algorithm which, with high probability, provides an almost uniform sample from the set of proper $k$-colourings on an instance of sparse random graphs $G_{n,d/n}$, where $k = k(d)$ is a sufficiently large constant. Our algorithm is not based on the Markov Chain Monte Carlo method (M.C.M.C.). Instead, we provide a novel proof of correctness of our Algorithm that is based on interesting "spatial mixing" properties of colourings of $G_{n,d/n}$. Our result improves upon previous results (based on M.C.M.C.) that required a number of colours growing unboundedly with $n$.

## 1 Introduction.

For a graph $G = (V, E)$, a (proper) $k$-colouring is an assignment $\sigma : V \to [k]$ such that adjacent vertices receive different colours, where for some positive integer $k$, $[k]$ indicates the set $\{1, \ldots, k\}$. It is well known that it is NP-hard to estimate the minimum number of colours in a proper $k$-colouring, i.e. estimate the chromatic number of $G$. However, in many cases there are estimates and upper bounds of the chromatic number e.g. if $\Delta$ is the maximum degree of $G$, then one can $k$-colour $G$ for $k = \Delta + 1$. Furthermore, for special classes of graphs the chromatic number has been estimated very accurately, e.g. in [1], Achlioptas and Naor have found the two possible values of the chromatic number that an instance of a sparse random graph has with high probability, i.e. with probability that tends to 1 as the size of the graph tends to infinity. All these raise the interesting computational challenge of finding the number of proper $k$-colourings for $k$ greater than the chromatic number.

In [14], Valiant, introduced the notion of $\#P$-hardness and proved that counting $k$-colourings is $\#P$-complete. The existence of a polynomial-time algorithm for exact counting is considered highly unlikely. Thus, we focus on designing polynomial-time algorithms for *approximate counting*. Practically, the closer $k$ gets to the chromatic number of $G$, the more difficult it becomes to estimate the number of its $k$-colourings. By [7], [8] we can reduce the *estimation of the number* of $k$-colourings of $G$ to *sampling* almost uniformly from the set of all its proper $k$-colourings. By "almost" we mean with distribution close, in some sense, to the uniform distribution.

In this work, we use focus on sampling $k$-colourings of instances of a sparse random graph, i.e. random graphs with vertices having expected degree equal to some constant, $d$, and $k = k(d)$ is a sufficiently large constant which scales as $d^{14}$, for sufficiently large $d$.

---

[*]Research Academic Computer Technology Institute, N. Kazantzaki str. Rio Patras, 26500, Greece and Computer Engineering and Informatics Department of the University of Patras, 26500, Greece. E-mail: euthimio@ceid.upatras.gr. CORRESPONDING AUTHOR

[†]Research Academic Computer Technology Institute N. Kazantzaki str. Rio Patras, 26500, Greece and Computer Engineering and Iinformatics Department of University of Patras, 26500, Greece. E-mail: spirakis@cti.gr

**Definition A** *Let $n$ be a positive integer and $p$, $0 \leq p \leq 1$. The random graph $G_{n,p}$ is a probability space over the set of graphs on the vertex set $\{1, \ldots, n\}$ determined by*

$$Pr[\{i,j\} \text{ is an edge of } G] = p$$

*with these events being mutually independent.*

For a sparse random graph the parameter $p$ is of the form $p = d/n$, where $d$ is a positive real constant. We take $d > 1$ (otherwise the problem of sampling is trivial).

The mathematical tool that we use for studying the problem of sampling $k$-colourings of instances of a sparse random graph is the "spin systems" and more specifically the proper colouring model, also refereed to as *antiferromagnetic Potts model at zero temperature* in statistical physics.

**Colouring Model on a finite graph.** The colouring model on a finite graph $G = (V, E)$ and set of colours $[\mathcal{S}]$, for some positive integer $\mathcal{S}$, is defined as follows. The system consists of a set of *sites*, which correspond to the vertices of $G$, and each site is assigned a *spin*, i.e. a member of $[\mathcal{S}]$. A *configuration* is an assignment of spins to $V$. Not all configurations can occur in the colouring model. A configuration that may occur is called a *feasible configuration*. The set of feasible configurations is the set of proper $\mathcal{S}$-colourings of the underlying graph $G$.

For any vertex set $V' \subseteq V$, let $\partial V' = \{v \in V \setminus V' | \exists u \in V' \text{ s.t. } \{u,v\} \in E\}$. Consider the colouring $C(\partial V') \in [\mathcal{S}]^{\partial V'}$, which is such that, there is a proper colouring in $[\mathcal{S}]^V$ with the vertices in $\partial V'$ coloured as $C(\partial V')$. In a system where the vertices in $\partial V'$ are coloured as $C(\partial V')$, the colour assignments of the vertices in $V'$ is distributed uniformly over, all the colour assignments of the vertices in $V' \cup \partial V'$ that agree with $C(\partial V')$ on the vertices in $\partial V'$ [1].

Frequently, one imposes *boundary conditions* on the model, which corresponds to fixing the colour assignment at some "*boundary*" vertex set of $G$; we use the term "*free boundary*" when there are no boundary conditions specified. In this work, when we consider a system with boundary conditions, we always assume that there exists at least one feasible configuration available for this system.

The probability to find a colouring model at a specific configuration is the uniform distribution over all proper colourings of the underlying graph. Generally, the probability of finding a system in a specific configuration is given by its *Gibbs* measure specified by this system.

**Definition B** *For a finite graph $G = (V, E)$ and an integer $\mathcal{S}$, let $PCS(G, \mathcal{S}, C(L))$ be a colourings model with underlying graph $G$, feasible configurations all the proper $\mathcal{S}$-colourings of $G$ and the boundary $L \subseteq V$ is coloured as $C(L)$. Let $\Omega(G, \mathcal{S}, C(L))$ be the set of feasible configurations of the system.*

The omission of the boundary conditions parameter implies free boundary. If the first parameter is a class of random graphs, e.g. $G_{n,p}$, then we consider that the underlying graph is an instance of this class.

Clearly, $\Omega(G, \mathcal{S}, C(L))$ is the set of all proper $\mathcal{S}$-colourings of $G$ that have the vertices in the set $L \subseteq V$ coloured as specified by the assignment $C(L)$. For a system $PCS(G, \mathcal{S}, C(L))$ we always assume that the boundary $C(L)$ is such that $\Omega(G, \mathcal{S}, C(L)) \neq \emptyset$.

For convenience, we use the following notation rules throughout this work. In $PCS(G = (V, E), \mathcal{S}, C(\Lambda))$, for $\Lambda \subseteq V$ the colour assignment of each vertex $v \in V$, or the set of vertices $V' \subseteq V$ are considered to be equal to the random variables $X_v^{C_\Lambda} \in [\mathcal{S}]$ and $X_{V'}^{C_\Lambda} \in [\mathcal{S}]^{V'}$, correspondingly.

---

[1] A rigorous definition of a colouring model involves the definition of a set of functions, the *compatibility functions* (see [16]). However, the definition we give here is a direct consequence of that with the compatibility functions.

**Definition C** *For the system $PCS(G = (V, E), S)$, the function $\mu(\cdot) : 2^{[S]^V} \to [0, 1]$ indicates the Gibbs measure specified by this system.*

In the system $PCS(G = (V, E), S, C(L))$, for $\forall v \in V$, we denote with $\mu(X_v | C(L))$, the marginal Gibbs measure of the random variable $X_v$.

## 1.1 Our work and related work.

**Previous work.** The pioneering work of Dyer et.al., in [4], proposes a very interesting Markov Chain Monte Carlo (MCMC) based algorithm, which with high probability (w.h.p.), i.e. with probability that tends to 1 as the size of graph tends to infinity, provides an almost uniform sample from the set of proper colourings of $G_{n,d/n}$ which uses at least $\Theta(\frac{\log \log n}{\log \log \log n})$ colours. Noting that, w.h.p. the maximum degree of a sparse random graph is $\Theta(\log n / \log \log n)$, to our knowledge, this work was the first to present a procedure for sampling colourings that uses fewer colours than the maximum degree.

**In parallel and independently, E. Mossel and A. Sly, have recently derived essentially the same result as we do here, (i.e. a random sampling $k$-colourings of a sparse random graph where $k$ is a constant) by using an MCMC approach, [12].**

**Our work.** Our approach is quite different. It is based on showing the validity of a specific *spatial mixing property* for the system $PCS(G_{n,d/n}, S)$, I.e. we show that if $S$ is greater than a specific value which depends *only* on the expected degree $d$, then an *asymptotic independence* between the colour assignment of any vertex $v$ and the colour assignment of any subset of vertices which is at sufficiently large (graph) distance from $v$ holds, in the system. We, then, present an algorithm which exploits this kind of asymptotic independence and produces, efficiently, an almost uniform sample of the $S$-colourings of the underlying graph $G_{n,d/n}$, which uses a **constant** number of colours (where the constant is an increasing function of $d$). For an earlier version of our result, see [5]. Our algorithm exploits ideas which are similar to those presented in [11] and [15], for counting satisfiable truth assignments in a random $k$-SAT formula and sampling independent sets of general graphs, correspondingly. However, our proof techniques are novel and some technical results are of independent interest.

A possible schema of our algorithm is the following one. The input is the finite graph $G = (V, E)$, an instance of $G_{n,d/n}$, and an integer $S$. We consider the system $PCS(G, S)$ and the algorithm provides a sample which is distributed "close" to the Gibbs measure specified by this system. The algorithm assumes an arbitrary permutation of the vertices of the input graph, e.g. $(v_1, v_2, \ldots, v_n)$, and in turn it assigns them a colour as follows: For $1 \le i \le n$, let $A_i \subseteq V$ be the set of $i - 1$ first coloured vertices, by the algorithm, and let $C(A_i)$ be their colour assignment. Assume that the colour assignment $C(A_i)$, of the vertices in $A_i$, is done according to a probability measure which is sufficiently close to $\mu(X_{A_i} = C(A_i))$, with $\mu(\cdot)$ beeing the Gibbs measure that is specified by $PCS(G, S)$. The algorithm computes efficiently a "good" estimation of $\mu(X_{v_i} = s | C(A_i))$, $\forall s \in [S]$, and assign $v_i$ a colouring according to this probability measure. The notion of "good" estimation of $\mu(X_{v_i} = s | C(A_i))$, $\forall s \in [S]$, implies that this estimation should be so accurate, that it will be possible for the algorithm to colour the remaining vertices with distribution as close to Gibbs measure of $PCS(G, S)$ as we initially wanted.

The spatial mixing property of $PCS(G_{n,d/n}, S)$, for sufficiently large $S$, is exploited by our algorithm in conjunction with the structural property of $G_{n,d/n}$ stated in Lemma A.

**Lemma A** *Let $G = (V, E)$ be an instance of $G_{n,d/n}$, where $d \ge 1$ is a fixed positive real. With high probability (w.h.p.) the graph has* no *vertex $v$ with the following property: The induced* subgraph of $G$ that contains $v$ and all vertices within graph distance $\epsilon \log n$ from $v$, contains more than one cycle, for any real $\epsilon > 0$ such that $\epsilon \leq (4 \log(e^2 d/2))^{-1}$.

The proof of Lemma A is given in section 2.1.

Showing an asymptotic independence between the colour assignment of any vertex $v$ and the colour assignment of any vertex set at distance greater than $\epsilon \log n$, for sufficiently small $\epsilon = \epsilon(d)$, implies that when the algorithm has to assign a colour to the $i$-th vertex the following holds: If the colouring $C(A_i)$ is done with probability measure which is sufficiently close to $\mu(X_{A_i} = C(A_i))$, then the algorithm can have a "good" estimation of $\mu(X_{v_i}|C(A_i))$ by just checking the colour assignments of a very simple structured neighborhood of $v_i$. The notion of "good" estimation is the same as the one stated previously. This kind of structure in the neighborhood of the vertex $v_i$ is highly desirable since then, it allows us to get a colouring of $v_i$, which is distributed as this "good" estimation of $\mu(X_{v_i}|C(A_i))$, in time which is upper bounded by a polynomial of $n$.

## 1.2 Further Definitions (Spatial Dependency)

For the graph $G = (V, E)$ and any two vertex sets $V', V'' \subset V$ we denote by $dist(V', V'')$ the graph distance of the two sets, i.e. the minimum length shortest path between all the pairs of vertices $(v_1, v_2) \in V' \times V''$.

**Definition D** *Let $G = (V, E)$ be an instance of $G_{n,d/n}$ and let $l$ be a postive real. For the vertex $v \in V$, let $G_{v,d,l}$ be the induced subgraph of $G$, which contains the vertex $v$ and all the vertices within graph distance $\lfloor l \rfloor$ from $v$.*

For a measure of comparison between probability measures we use the *total variation distance*.

**Definition E** *For measures $\mu$ and $\nu$ on the same discrete space $\Omega$, the total variation distance $d_{TV}(\mu, \nu)$ between $\mu$ and $\nu$ is defined as*

$$d_{TV}(\mu, \nu) = \frac{1}{2} \sum_{x \in \Omega} |\mu(x) - \nu(x)|$$

**Definition F (Spatial Dependency.)** *Consider the graph $G = (V, E)$, an instance of $G_{n,d/n}$, the positive integers $\mathcal{S}$, $l$ and the positive real $s$. For each $v \in V$ consider the subgraph $G_{v,d,l}$ with vertex set $V_{v,l}$. For a given $v \in V$, consider also any two $\mathcal{S}$-colourings $C_1(V_1)$ and $C_2(V_1)$ with $V_1 \subset V_{v,l}$ and $dist(\{v\}, V_1) \geq l$, having $\Omega(G_{v,l}, \mathcal{S}, C_1(V_1))$, $\Omega(G_{v,l}, \mathcal{S}, C_2(V_1)) \neq \emptyset$. If for $\forall v \in V$ it holds*

$$d_{TV}\left(\tilde{\mu}(X_v|C_1(V_1)), \tilde{\mu}(X_v|C_2(V_1))\right) \leq s$$

*with $\tilde{\mu}(\cdot)$ specified by the system $PCS(G_{v,d,l}, \mathcal{S})$, then we say that " $\forall v$ the distance $l$ **Spatial Dependency** of the $\mathcal{S}$-colourings of $G$ is $s$". This will be denoted as $\forall v \in V \; SD(v, l) = s$.*

Clearly, the above definition extends, directly, to any system with underlying graph any type of graph.

It is easy to see that if "$\forall v \in V$ the distance $l$ Spatial Dependency of the $\mathcal{S}$-colourings of $G = (V, E)$ is $s$", then in the system $PCS(G, \mathcal{S})$ and for each vertex $v \in V$ any two colourings $C_1'(V_1)$ and $C_2'(V_1)$ with $V_1$ as defined in Definition F and $C_1'$ and $C_2'$ having $\Omega(G, \mathcal{S}, C_1'(V_1)) \neq \emptyset$ and $\Omega(G, \mathcal{S}, C_2'(V_1)) \neq \emptyset$ it holds

$$d_{TV}\left(\mu(X_v|C_1(V_1)), \mu(X_v|C_2(V_1))\right) \leq s$$

with $\mu(\cdot)$ specified, now, by the system $PCS(G, \mathcal{S})$. The above holds since for each $v \in V$, if $\Omega_1 = \{C(V_1)|\Omega(G_{v,l}, \mathcal{S}, C(V_1)) \neq \emptyset\}$ and $\Omega_2 = \{C(V_1)|\Omega(G, \mathcal{S}, C(V_1)) \neq \emptyset\}$, then clearly $\Omega_2 \subseteq \Omega_1$.

## 1.3 Structure of the remaining paper.

The remaining of our paper has the following structure. In section 2, we present two basic properties of the spin systems which we deal with. Then, it follows a detail description of our sampling algorithm in a form of pseudo code accompanied by a discussion and a statement, without proof, of two theorems which deal with the accuracy of the result that is returned and the execution time of the algorithm, corespondingly.

The proofs and an analytic discussion about the properties of the spin-system that our algorithm considers, are given in section 3. This work ends by presenting the proofs of the theorems, that deal with the accuracy and the time efficiency of the our sampling algorithm, section 4.

For easiness of verification of our proofs, we provide here the dependence of each lemma and theorem on lemmas logically preceding it: Lemma A, Lemma B and Lemma C, have no lemmas preceding them. Lemma D has predecessors Lemma A, Lemma B and Lemma C. Lemma E has predecessors Lemma A, Lemma B, Lemma C and Lemma E. Lemma F has no predecessors. Lemma G has predecessors Lemma C and Lemma B. Lemma H has predecessors, Lemma F and Lemma G.

Theorem A has precedings all the lemmas of this paper. Theorem B has predecessors, Theorem A, Lemma A. Theorem C, has predecessors, Theorem A, Theorem B and Lemma A. Finally, Theorem D has no preceding.

## 2 Statement of results.

### 2.1 Properties of the spin system.

In this section we state two crucial properties that the colouring model has with high probability, when the underlying graph is an instance of a sparse random graph. These properties are stated in the following lemma and theorem. The first one, already stated in section 1, refers to the structure of the neighborhood of each vertex in an instance of a sparse random graph. The second one refers to a property of the colourings (configurations) of such a spin-system. I.e. in a $PCS(G_{n,d/n}, \mathcal{S})$, if $\mathcal{S}$ is greater than a specific value which depends *only* on the expected degree $d$, then an *asymptotic independence* between the colour assignment of any vertex $v$ and the colour assignment of any subset of vertices which is at graph distance, at least, $\left\lfloor \frac{0.9}{4\log(e^2 d/2)} \log n \right\rfloor$ from $v$ holds.

**Lemma A** *Let $G = (V, E)$ be an instance of $G_{n,d/n}$, where $d \geq 1$ is a fixed positive real. With high probability (w.h.p.) the graph has no vertex $v$ with the following property: The induced subgraph of $G$ that contains $v$ and all vertices within graph distance $\epsilon \log n$ from $v$, contains more than one cycle, for any real $\epsilon > 0$ such that $\epsilon \leq (4\log(e^2 d/2))^{-1}$.*

**Proof:**
To show the lemma assume the contrary, i.e. there is some vertex $v \in V$ whose corresponding graph $G_{v,d,\epsilon \log n}$ (see Definition D) contains two cycles, i.e. $C_1$ and $C_2$ each of length at most $2\epsilon \log n$, the value of $\epsilon$ will be determined later. The above assumption implies that there are two pairs of paths starting from $v$, such that: The paths in each pair do not have all their edges common and there is some vertex in $G_{v,d,\epsilon \log n}$ that can be reached from $v$ by both paths of the pair. The existence of such two pairs of paths implies that in $G_{v,d,\epsilon \log n}$ there is a set of, at most $4\epsilon \log n$, vertices which have among each other a number of edges which exceeds the number of vertices by one.

Thus, the proof of the lemma reduces to showing that in $G_{n,d/n}$, there is no set of, at most, $4\epsilon \log n$, vertices which contains a number of edges that exceed the number of vertices in the set by one, for sufficiently small $\epsilon$. Let $D$ be the event "such a set exists". Setting $r = 4\epsilon \log n$ we have

$$\begin{aligned} Pr[D] &\leq \sum_{k=1}^{r} \binom{n}{k} \binom{\binom{k}{2}}{k+1} \left(\frac{d}{n}\right)^{(k+1)} \\ &\leq \sum_{k=1}^{r} \left(\frac{ne}{k}\right)^{k} \left(\frac{ek(k+1)}{2(k+1)}\right)^{(k+1)} \left(\frac{d}{n}\right)^{(k+1)} \\ &\leq \frac{de}{2n} \sum_{k=1}^{r} \left(\frac{e^2}{2}\right)^{k} kd^k \\ &\leq \frac{d^2 e^3}{4} \frac{4\epsilon \log n}{n} \sum_{k=0}^{r-1} \left(\frac{de^2}{2}\right)^{k} \\ &\leq \frac{d^2 e^3 \epsilon \log n}{n} \frac{(de^2/2)^r - 1}{de^2/2 - 1} \end{aligned}$$

taking $\epsilon$ such that $4\epsilon \cdot \log(de^2/2) < 1$ the r.h.s. of the last equation is $o(1)$, as the nominator is $o(n)$. Thus, for sufficiently small $\epsilon$ we have that $Pr[D] = o(1)$ ◇

**Theorem A** *Let $G = (V, E)$ be an instance of $G_{n,d/n}$, where $d > 1$. If $\mathcal{S}$ is a sufficiently large integer, which depends on d, and $\epsilon = \frac{0.9}{4\log(e^2 d/2)}$, then w.h.p., i.e. with probability $1 - 2n^{-0.25}$, for every vertex $v \in V$ the distance $\lfloor \epsilon \log n \rfloor$ Spatial Dependency of the $\mathcal{S}$-colourings of $PCS(G, \mathcal{S})$ is $n^{-1.25}$. For sufficiently large d, we should have $\mathcal{S} \geq d^{14}$.*

We mention that, if $d$ is relatively small, then for $PCS(G_{n,d/n}, \mathcal{S})$, the distance $\lfloor \epsilon \log n \rfloor$ Spatial Dependency of the $\mathcal{S}$-colourings of $PCS(G, \mathcal{S})$ can become $n^{-1.25}$, however, for a number of colours which is a constant greater than $d^{14}$.

Section 3 is devoted to the proof of Theorem A.

## 2.2 Our Algorithm

We start by presenting our sampling algorithm in the form of pseudo code. The input of the algorithm is the graph $G = (V, E)$ an instance of $G_{n,d/n}$ with $d > 1$, and the integer $\mathcal{S}$. Conditioning on the properties of the system stated in the previous section, which hold w.h.p. by taking sufficiently large constant $\mathcal{S}$, the algorithm outputs a $\mathcal{S}$-colouring of the input graph distributed within total variation distance $n^{-0.25}$ from the uniform distribution over the set of all $\mathcal{S}$-colourings of the input graph.

We have to mention, here, that our algorithm is based on properties of the spin-system that already hold w.h.p.. This means that we can expose the entire instance of the input graph at the beginning and expect these, desired, properties to hold, which is highly likely.

In what follows, we assume that $v_i$ is the $i$-th vertex to be coloured by the algorithm [2] and $A_i$ is the set of vertices that have already been coloured, before $v_i$. We, also, assume that $A_i$ is coloured as $C(A_i)$. We denote with $(V_i, E_i)$ the vertex set and the edge set, correspondingly, of the graph $G_{v_i, d, \epsilon \log n}$, where $\epsilon = \frac{0.9}{4\log(de^2/2)}$

---
[2] We remind the reader that the algorithm colours one vertex the time.

**Sampling Algorithm**
    **Input:** $G = (V, E)$, instance of $G_{n,d/n}$, number of colours $\mathcal{S}$
    Take an arbitrary permutation of the vertices in $V$, i.e. $(v_1, \ldots, v_n)$
    $A_1 = \emptyset$
    **For** $i = 1, \ldots, n$
        -Create the subgraph $G_{v_i, d, \epsilon \log n} = (V_i, E_i)$
        -**If** $G_{v_i, d, \epsilon \log n} = (V_i, E_i)$ is not a tree or a unicyclic graph
          **Then Return** Failure
        -Colour $v_i$ according to $\tilde{\mu}_i(X_{v_i} | C(A_i \cap V_i))$
        using dynamic programming
        -$A_{i+1} := A_i \cup \{v_i\}$
    **Return** Colouring of $G$

The probability measure $\tilde{\mu}_i(\cdot)$ is the Gibbs measure specified by the system $PCS(G_{v_i, d, \epsilon \log n}, \mathcal{S})$.

Note that if the algorithm, at each iteration of the for loop, had assigned to the vertex $v_i$ a colouring, according to $\mu(X_{v_i} | C(A_i))$, instead of $\tilde{\mu}_i(X_{v_i} | C(A_i \cap V_i))$, then it would have been exact. I.e., the distribution $\tilde{\mu}_i(X_{v_i} | C(A_i \cap V_i))$ is an estimation of $\mu(X_{v_i} | C(A_i))$ for our algorithm.

There are two issues to be clarified about the algorithm above. The first one is its **accuracy**, i.e. how close is the distribution of the colouring that is returned, to the uniform distribution over all proper $\mathcal{S}$-colourings of the input graph. The second one is its **efficiency**, i.e. how much time is needed for the execution of the algorithm with respect to the size of the input graph.

As far as the accuracy of the algorithm is regarded, we use Theorem A. By the discussion at the end of section 1.2, we see that Theorem A implies that, w.h.p. the Gibbs measure specified by $PCS(G_{n,d/n}, \mathcal{S})$, where $\mathcal{S}$ is a sufficiently large constant, is such that an *asymptotic independence* (spatial mixing) holds, for the colour assignment of any vertex $v$ and the colour assignment of any vertex set at distance $\lfloor \epsilon \log n \rfloor$ from $v$, where $\epsilon = \frac{0.9}{4 \log(e^2 d/2)}$.

We claim that at the $i$-th iteration of the for loop of the algorithm, the following holds: If the algorithm has assigned a colouring to the vertices in $A_i$, according to the Gibbs measure, $\mu(\cdot)$, that specifies $PCS(G, \mathcal{S})$, then it, still, holds the same asymptotic independence as that it is implied by Theorem A, between the colouring of the vertex $v_i$ and the colourings of vertices at distance, at least, $\lfloor \epsilon \log n \rfloor$ from $v_i$, where $\epsilon = \frac{0.9}{4 \log(e^2 d/2)}$.

Equivalently, we can think of the following situation. Consider two systems, i.e. $S_1 = PCS(G, \mathcal{S})$ and $S_2 = PCS(G, \mathcal{S})$, with underlying graph $G$, the input of our algorithm and with each system being independent of the other. Assume that, in both systems, we fix the colour assignments of the vertices in $A_i \subset V$ according to $\mu(X_{A_i})$, with $\mu(\cdot)$ specified by $PCS(G, \mathcal{S})$. Assume that after having fixed the colour assignments, of the vertices in $A_i$, we look at the colour assignments of the vertices at graph distance, at least, $\lfloor \epsilon \log n \rfloor$ from the vertex $v_i$, in both systems. Let $V'$ be the vertex set whose colourings have been seen and let $C(V')$ be the colouring we see in $S_1$ and $C'(V')$ be the colouring we see in $S_2$. The above claim, is equivalent to saying that knowing the colour assignments of the vertices in $V'$ in both systems, then the total variation distance between the probability measures of the colour assignments of $v_i$ in the two systems, correspondingly, is upper bounded by the quanitity $SD(v_i, \lfloor \epsilon \log n \rfloor)$. I.e.

$$d_{TV}\left(E[\mu(X_{v_i} | X_{A_i}, X_{V'} = C(V'))], E[\tilde{\mu}_i(X_{v_i} | X_{A_i}, X_{V'} = C'(V'))]\right) \leq n^{-1.25}. \tag{1}$$

Where both expectations, are taken over all colourings of $X_{A_i}$ where the probability of each colouring is according to Gibbs measure $\mu(\cdot)$.

The following theorem, Theorem B, (in the proof of which it is shown the validity of the claim above) gives a characterization of the distribution of the colouring that is returned by the algorithm in terms of its total variation distance from the uniform over all the proper $\mathcal{S}$-colourings of the input graph, if $\mathcal{S}$ is as large as indicated in Theorem A.

**Theorem B** *If $\mathcal{S}$ is a sufficiently large integer constant, then, with probability $1 - O(n^{-0.1})$, the sampling algorithm is successful and returns a $\mathcal{S}$-colouring of the input graph $G$, whose distribution is within total variation distance $n^{-0.25}$ from the uniform over all the proper $\mathcal{S}$-colourings of $G$.*

The proof of Theorem B is given in section 4.
As far as the execution time of the algorithm is concerned, we make the following remark. According to Lemma A, the set $\{G_{v_i,d,\epsilon \log n}, \text{ for } i = 1, \ldots, n\}$, for $\epsilon = \frac{0.9}{4 \log(e^2 d/2)}$, w.h.p., i.e. with probability $1 - n^{-0.1}$, contains graphs which are either unicyclic or trees. If this is not the case, then we consider that the algorithm fails. As argued in [4], we can have a colouring of the vertex $v_i$ according to $\tilde{\mu}_i(X_{v_i}|C(A_i \cap V_i))$ by generating a random colouring of $G_{v_i,d,\epsilon \log n}$ where the vertices in $A_i \cap V_i$ are colored as $C(A_i \cap V_i)$ in time upper bounded by $l \cdot k^3$, where $l = |V_i|$ and $k = \mathcal{S}$ (for more details see the proof of Theorem C and [4]).

**Theorem C** *The time complexity of the sampling algorithm is w.h.p. asymptotically upper bounded by $O(n^2)$, where $n$ is the number of vertices of the input graph.*

The proof of Theorem C is given in section 4.
We note that at the $i$-th iteration of the for-loop of the algorithm, we can apply the Junction tree algorithm (see [16]) to assign the vertex $v_i$ a colour according to the probability measure $\tilde{\mu}_i(X_u|C(A_i \cap V_i))$. The execution time of the junction tree is asymptotically bounded by $O(n^{2+c})$, where $c < 1$ is a sufficiently large constant.

## 3 Spatial mixing.

According to Lemma A, if $G = (V, E)$ is an instance of $G_{n,d/n}$, then the set of graphs $G_{v,d,\epsilon \log n}$, for $v \in V$, with $\epsilon = \frac{0.9}{\log(e^2 d/2)}$, w.h.p. contains graphs which, each of them, is either unicyclic or trees. Instead of prooving Theorem A, equivalently, we show the two following lemmas, which are proved in sections 3.2 and 3.3, correspondingly.

**Lemma E** *Consider the system $PCS(G_{v,d,\epsilon \log n}, \mathcal{S})$, for $d > 1$, $\epsilon = \frac{0.9}{4 \log(e^2 d/2)}$ and for $G_{v,d,\epsilon \log n}$ we condition that it is a tree. If $\mathcal{S}$ is a sufficiently large constant, then with probability at least $1 - 2n^{-1.25}$, for the above system it holds that $SD(v, \lfloor \epsilon \log n \rfloor) = n^{-1.25}$. For sufficiently large $d$, we should have $\mathcal{S} \geq d^{14}$.*

**Lemma H** *Consider the system $PCS(G_{v,d,\epsilon \log n}, \mathcal{S})$, for $d > 1$, $\epsilon = \frac{0.9}{4 \log(e^2 d/2)}$ and for $G_{v,d,\epsilon \log n}$ we condition that it is a unicyclic graph. If $\mathcal{S}$ is a sufficiently large constant, then with probability at least $1 - 2n^{-1.25}$, for the above system it holds that $SD(v, \lfloor \epsilon \log n \rfloor) = n^{-1.25}$. For sufficiently large $d$, we should have $\mathcal{S} \geq d^{14}$.*

One can see that the lemmas E and H imply Theorem A, see Corollary A.

**Corollary A** *If Lemma E and Lemma H are true, then Theorem A is true, as well.*

**Proof:** Assume that Lemma E and Lemmas H are true. Consider the system $PCS(G_{n,d/n}, \mathcal{S})$, where $\mathcal{S}$ is a sufficiently large constant and for sufficiently large $d$, $\mathcal{S} \geq d^{14}$. Theorem A holds for $PCS(G_{n,d/n}, \mathcal{S})$ if the following event holds with probability at least $1 - 2n^{-0.25}$

$$Event_1 = \text{"}\textit{for every} \text{ graph } G_{v,d,\epsilon \log n} \text{ of the set of graphs that } G_{n,d/n} \text{ specifies, it holds}$$
$$\text{that the } PCS(G_{v,d,\epsilon \log n}, \mathcal{S}) \text{ has the property that } SD(v, \lfloor \epsilon \log n \rfloor) = n^{-1.25}\text{"}$$

By Lemma E and Lemma H we have that for sufficiently large $\mathcal{S}$, which for sufficiently large $d$ becomes $\mathcal{S} \geq d^{14}$, such that, for every vertex $v$ in $G_{n,d/n}$ the

$$Event_v = \text{" the system } PCS(G_{v,d,\epsilon \log n}, \mathcal{S}) \text{ has the property that } SD(v, \lfloor \epsilon \log n \rfloor) = n^{-1.25}\text{"}$$

holds with probability at least $1 - 2n^{-1.25}$. Clearly

$$Pr[Event_1] = 1 - Pr\left[\cup_v \overline{Event_v}\right]$$

By the union bound we get that $Pr[Event_1] \geq 1 - 2n^{-0.25}$, which proves corollary. ◇

**Note.** In both cases the crucial point is to show that a certain quantity (the "disagreement probability") has a small expected value. This quantity measures how much we deviate if, for colouring some vertex $v$ in the algorithm, we consider only a heighborhood of its, instead of the whole graph. In the sequel we carefully establish the necessary upper bounds for this.

**Remark.** Both Lemma E and Lemma H are based on the fact that we expect a very large proportion of the vertices of an instance of $G_{n,d/n}$ to have constant degrees. I.e. there is a constant $c_0 = c_0(d)$ such that for any $c > c_0$ the expected proportion of vertices that have degree less than $c$ tends to 1, exponentially fast with $c$. This argument is justified by the following corollary that is proved in [6].

**Corollary B** *If a random variable $Z$ is distributed as in $B(n, q)$, the binomial distribution with parameters $n$ and $q$, with $\lambda = nq$ then*

$$Pr[Z \geq x] \leq e^{-x} \quad x \geq 7\lambda.$$

### 3.1 The process ColourRoot and a coupling.

Towards proving Lemma E and Lemma H, we introduce, here, the stochastic process *ColourRoot* $(T, \mathcal{S}, C(L))$, where $T = (V, E)$ is a tree, $\mathcal{S}$ is a positive integer and the vertices in $L \subset V$ are assigned a colouring $C(L)$, such that $\Omega(T, \mathcal{S}, C(L)) \neq \emptyset$. The process *ColourRoot*$(T, S, C(L))$ assigns a colouring (not necessarily proper), to the vertices in $V \setminus L$ such that $\forall u \in V \setminus L$ its colour assignment is distributed as in $\mu(X_u | C(L \cap T_u))$, where $T_u$ is the subtree of $T$ rooted at $u$ while the Gibbs measure $\mu(X_v | C(L \cap T_u))$ is specified by the system $PCS(T_u, \mathcal{S}, C(L \cap T_u))$. When the third parameter of the ColourRoot is omitted, it is implied that there is no fixed colour assignment to any vertex.

The ColourRoot$(T, \mathcal{S}, C(L))$ assigns a colouring to each vertex $u$ in the tree $T$, based on the following observation. For the vertex $u$ of $T$ consider the vertex set $CH_u$ which contains the children of $u$ in $T_u$ and the system $PCS(T_u, \mathcal{S}, C(L \cap T_u))$. For the graph $T_0 = \cup_{w \in CH_u} T_w$ consider the set of $\mathcal{S}$-colorings $\Omega_0 = \Omega(T_0, \mathcal{S}, C(L \cap T_0))$ Assume that each $C \in \Omega_0$ specifies a colouring of the vertices in $CH_u$ that uses all but $W_C$ colours from the set $[\mathcal{S}]$. Note that if the system $PCS(T_u, \mathcal{S}, C(L \cap T_u))$ is in equilibrium, the probability for the vertices in $T_0 = \cup_{w \in CH_u} T_w$ to be coloured as specified by $C \in \Omega_0$ is proportional to the quantity $W_C$, i.e. $\frac{W_C}{\sum_{C \in \Omega_0} W_C}$.

**Definition G** *With the above notation, the process ColourRoot(T, $\mathcal{S}$, C(L)) assigns a colour to the vertex u of T, as follows:*

1. *Each $C \in \Omega_0$ is assigned weight, $W_C$, equal to the number of colours in the set $[\mathcal{S}]$ that do not appear in the colour assignment that C specifies for the vertices in $CH_u$.*

2. *Select from $\Omega_0$ such that the probability for each member to be chosen is proportional to the weight it has been assigned to it. Let $C'$ be the chosen member.*

3. *Assign to the vertex u a colour that is chosen uniformly at random among the colours in the set $[\mathcal{S}]$ that do not appear in the colouring of the vertices in $CH_u$, as this is specified by $C'$.*

For a *coupling* of the processes ColourRoot$(T, \mathcal{S}, C(L))$ and ColourRoot $(T, \mathcal{S}, C'(L))$, we introduce the notion of *disagreement probability*.

**Definition H** *Consider a coupling of ColourRoot $(T, \mathcal{S}, C(L))$ and ColourRoot$(T, \mathcal{S}, C'(L))$. The disagreement probability for a vertex u in T, denoted by $p_u$, is equal to the probability for the coupling to assign different colours to u.*

The coupling of ColourRoot $(T, \mathcal{S}, C(L))$ and ColourRoot$(T, \mathcal{S}, C'(L))$ is of our main interest here, due to the following, very significant fact.

**Theorem D** *Consider the tree $T = (V, E)$ rooted at the vertex r, some set $A \subseteq V$ and any two colourings of the vertex set A, such that $\Omega(T, \mathcal{S}, C(A)), \Omega(T, \mathcal{S}, C'(A)) \neq \emptyset$. Assume that there is a coupling of the ColourRoot$(T, \mathcal{S}, C(A))$ and the ColourRoot$(T, \mathcal{S}, C'(A))$, for some integer $\mathcal{S}$, such that the probability of disagreement for the root r is $p_r$. Then, it holds that*

$$d_{TV}(\mu(X_r|C(A)), \mu(X_r|C'(A))) \leq p_r.$$

*where $\mu(X_r|C(A))$ and $\mu(X_r|C'(A))$ are specified by the system $PCS(T, \mathcal{S})$.*

**Proof:** The theorem follows directly from the Coupling Lemma (see [2]). $\diamond$

For the system $PCS(T, \mathcal{S})$, where $T$ is a tree rooted at vertex $r$, one can derive upper bounds for $SD(r, l)$, for some positive integer $l$, by using the above theorem and the coupling of the ColourRoot, as described in the following definition.

**Definition I** *Consider the tree $T = (V, E)$ rooted at vertex r, an integer $\mathcal{S}$ and the set $V_1 \subset V$ such that $dist(\{r\}, V_1) \geq l$ for some integer l. Let $\mathcal{C}(T, \mathcal{S}, l)$ be the coupling of the processes ColourRoot$(T, \mathcal{S}, C_1(V_1))$ and ColourRoot$(T, \mathcal{S})$. The colour assignment $C_1(V_1)$, is taken so as to* maximize *the disagreement probability at the root of T. The coupling $\mathcal{C}(T, \mathcal{S}, l)$ assigns colours to the vertex u of T as follows:*

- *Couple step 2 of the two processes so as to maximize the probability for the vertices in $CH_u$ to have the same colour assignment.*

- *Conditional on the choices the two processes have made at their step 2, assign colours to u, so as to minimize the disagreement probability $p_u$.*

In the coupling $\mathcal{C}(T, \mathcal{S}, l)$, if the height of $T$ is less than $l$, then the set $V_1$, as given in Definition I, is empty. It is easy for one to see that in that case, the disagreement probability for all vertices in $T$, is zero.

**Corollary C** *Consider a tree $T$, rooted at vertex $r$. If the coupling $\mathcal{C}(T,\mathcal{S},l)$, for some positive integers $\mathcal{S},l$, has disagreement probability $p_r$ for the root of $T$, then for the system $PCS(T,\mathcal{S})$ it holds that $SD(r,l) \leq 2p_r$.*

**Proof:** For the tree $T$, rooted at vertex $r$, and the integers $\mathcal{S}$, $l$, assume that the coupling $\mathcal{C}(T,\mathcal{S},l)$ has disagreement probability on the root $p_r$. Consider the vertex set $L$, which contains vertices at distance, at least, $l$ from the root $r$. Let, also, $\tilde{C}(L)$ and $\hat{C}(L)$ be the two colourings which maximize the total variation distance of the measures $\mu(X_r|C(\tilde{C}(L))$ and $\mu(X_r|\hat{C}(L))$, as these are specified by the system $PCS(T,\mathcal{S})$. It holds that

$$\begin{aligned} SD(r,l) &= d_{TV}\left(\mu(X_r|C(\tilde{C}(L)), \mu(X_r|\hat{C}(L))\right) \\ &\leq d_{TV}\left(\mu(X_r|C(\tilde{C}(L)), \mu(X_r)\right) + d_{TV}\left(\mu(X_r), \mu(X_r|\hat{C}(L))\right) \\ &\leq 2p_r \end{aligned}$$

the second direvation follows by the triangle inequality for measures. The corollary follows. $\diamond$

We note to the reader that, here, we will not need to give an explicit desciption of the coupling $\mathcal{C}(T,\mathcal{S},l)$. It suffices to show that $\mathcal{C}(T,\mathcal{S},l)$ has two specific properties, those that indicated by Lemma B and Lemma C[3].

In the rest of this section, we state and prove Lemma B and Lemma C. These two lemmas provide means to derive upper bounds for the probability of disagreement, in the coupling $\mathcal{C}(T,\mathcal{S},l)$, for each vertex $u$ of $T$. More specifically, Lemma B and Lemma C provide an inductive description of the coupling $\mathcal{C}(T,\mathcal{S},l)$, in terms of the disagreement probabilities. I.e. considering the vertex $u$ and the set $CH_u$ of its children, if the coupling $\mathcal{C}$ assigns colours to each vertex $w \in CH_u$ such that the probability of disagreement is $p_w$, then for the vertex $u$ the probability of disagreement $p_u$, in $\mathcal{C}$, can be bounded as follows

$$p_u \leq a(|CH_u|, \mathcal{S}) \cdot \left( \sum_{w \in CH_u} p_w \right)$$

where $a(|CH_u|, \mathcal{S})$ is a constant that depends on the cardinality of $CH_u$ and $\mathcal{S}$.

We distinguish two classes of vertices in $T$ regarding the relation between their number of children and the number of available colours $\mathcal{S}$, i.e. the *mixing* vertices and the *nonmixing* vertices. The mixing vertices have a number of children which is smaller than $\mathcal{S}$ and the constant $a(|CH_u|, \mathcal{S})$ is very small for them, i.e. $<< 1$. The *nonmixing* vertices have high degrees and for them the constant $a(|CH_u|, \mathcal{S})$ may become very large.

**Definition J** *Each vertex $u$ of the tree $T$ is "mixing" if, for a given $t$, the number of its children in $T$ is less than or equal to $t$, otherwise it is "nonmixing".*

The value of $t$, the maximum number of children of a mixing vertex, in the coupling $\mathcal{C}(T,\mathcal{S},l)$ is *always less than the number of available colours*. Generally, for a given tree $T$ and number of colours $\mathcal{S}$, we take $t$ so large as to minimize the probability of disagreement of the root of the tree $T$ in $\mathcal{C}(T,\mathcal{S},l)$.

---

[3] However, if the reader is keen on finding one, then he can deduce one from the proofs of Lemma B and Lemma C and the proofs of the claims inside them.

**Lemma B** *Let $u$ be a vertex of the tree $T$ which is* mixing. *If for every vertex $w \in CH_u$ the probability of disagreement, in the coupling $\mathcal{C}(T, \mathcal{S}, l)$, is $p_w$, then for the vertex $u$ the probability of disagreement $p_u$, in $\mathcal{C}$, is bounded as*

$$p_u \leq \frac{t \cdot \mathcal{S}}{(\mathcal{S}-t)^2} \cdot \left( \sum_{w \in CH_u} p_w \right)$$

*where $t$ is the maximum number of children of a mixing vertex.*

**Proof:** Assume that in the coupling $\mathcal{C}(T, \mathcal{S}, l)$, when the vertex $u$ is to be coloured, the processes ColourRoot$(T, \mathcal{S}, C_1(V_1))$ and ColourRoot$(T, \mathcal{S})$ at their second step choose from the set of colourings $\Omega_C$ and $\Omega_F$, correspondingly.

Let $\mathcal{A}$ be the event that, when the vertex $u$ is to be coloured in $\mathcal{C}$ the members of $\Omega_F$ and $\Omega_C$, that are chosen at step (2) of the processes ColourRoot, specify different colour assignments for the vertices in $CH_u$. Then

$$Pr[\text{disagreement on } u] = Pr[\text{disagreement on } u|\mathcal{A}]Pr[\mathcal{A}] + Pr[\text{disagreement on } u|\overline{\mathcal{A}}]Pr[\overline{\mathcal{A}}]$$

We mention that, if the event $\mathcal{A}$ does not hold ($\overline{\mathcal{A}}$ holds), then there is a coupling for step (3) of the ColourRoot that assigns the same colour to the vertex $u$ in $\mathcal{C}$, i.e. $Pr[\text{disagreement on } u|\overline{\mathcal{A}}] = 0$. Thus,

$$Pr[\text{disagreement on } u] = Pr[\text{disagreement on } u|\mathcal{A}]Pr[\mathcal{A}] \qquad (2)$$

To show the lemma we provide appropriate upper bounds for the probabilities in (2), i.e. $Pr[\mathcal{A}]$ and $Pr[\text{disagreement on } u|\mathcal{A}]$. We start with bounding $Pr[\mathcal{A}]$,

We note that the assumption that for each $w \in CH_u$, the disagreement probability in $\mathcal{C}$, is $p_w$ can be seen as follows: There is a coupling, call it $\mathcal{K}_1$, which chooses uniformly at random (u.a.r.) from the sets $\Omega_F$ and $\Omega_C$ and the two chosen elements specify different colour assignments for the vertices in $CH_u$ with probability which is upper bounded by $\sum_{w \in CH_u} p_w$.

Note that $|\Omega_F| \neq |\Omega_C|$. We create the set $\Omega'_F$ such that, each element of $\Omega_F$ appears $|\Omega_C|$ times in $\Omega'_F$. Similarly, we create the set $\Omega'_C$ such that, each element of $\Omega_C$ appears $|\Omega_F|$ times in $\Omega'_C$. Clearly, $|\Omega'_C| = |\Omega'_F|$.

**Claim A** *We can have a coupling, call it $\mathcal{K}_2$, that chooses uniformly at random (u.a.r.) an element from each of the sets $\Omega'_C$ and $\Omega'_F$, such that the probability for the two chosen elements to specify different colour assignment for any vertex in $CH_u$ is upper bounded by $\sum_{w \in CH_u} p_w$.*

Assume that in $\mathcal{C}$, each of the executions of ColourRoot, at step (2), considers the sets $\Omega'_C$ and $\Omega'_F$, correspondingly, instead of $\Omega_C$ and $\Omega_F$. Clearly, the fact that each of the processes ColourRoot considers the set $\Omega'_C$ instead of $\Omega_C$ and $\Omega'_F$ instead of $\Omega_F$, correspondingly, does not change the marginal probability measure of the colour assignment of the vertex $v$, in the coupling $\mathcal{C}$.

**Claim B** *Assume that the number of children of the vertex $u$ is $k$ and the disagreement probability for $w \in CH_u$ is $p_w$. If at the coupling of the second step of the processes ColourRoot in $\mathcal{C}(T, \mathcal{S}, l)$, each $C \in \Omega'_C \cup \Omega'_F$ is assigned weight $W_C$, then for the event $\mathcal{A}$ we have that*

$$Pr[\mathcal{A}] \leq \frac{1}{q_{k,\mathcal{S}}} \frac{max_{C \in \Omega'_F \cup \Omega'_C}\{W_C\}}{min_{C \in \Omega'_F \cup \Omega'_C}\{W_C | W_C > 0\}} \sum_{w \in CH_u} p_w.$$

*where $q_{k,\mathcal{S}}$ is the probability of the event that after $k$ trials, not all elements of $[\mathcal{S}]$ have been chosen, when at each trial we choose u.a.r. a member of $[\mathcal{S}]$.*

Note that the number of children of a mixing vertex is less than the number of available colours, thus, there is no colouring of the vertices in $CH_u$ that leave no available colour for $u$. Since we have assumed that $u$ is *mixing*, it is clear that in our case $q_{k,S} = 1$. Also note that at the coupling of step (2) of the ColourRoot, for colouring the vertex $u$, no member of either $\Omega'_C$ or $\Omega'_F$ is assigned a weight which is more than $S$ and less than $S - t$, where $t$ is equal to the maximum number of children that a mixing vertex can have. Thus, we have

$$Pr[\mathcal{A}] \leq \frac{S}{S - t} \sum_{w \in CH_u} p_w$$

We proceed to derive a bound for $Pr[\text{disagreement on } u|\mathcal{A}]$. For this we use the following claim.

**Claim C** *Consider the coupling $\mathcal{C}(T, S, l)$ when it assigns colourings to the vertex $u$. Assume that the two processes chose members of $\Omega'_F$ and $\Omega'_C$ that specify colourings for the vertices in $CH_u$ such that for the vertex $u$, there are two lists of available colours, $l_1$ and $l_2$, correspondingly. Assuming that $|l_i| > 0$, for $i = 1, 2$, there is a coupling that can choose the same colour for the vertex $u$ with probability at least*

$$1 - \frac{max\{|l_1 \setminus l_2|, |l_2 \setminus l_1|\}}{min\{|l_1|, |l_2|\}}$$

.

By Claim C we have that

$$Pr[\text{disagreement on } u|\mathcal{A}] \leq \frac{t}{|S| - t}$$

since, $|l_1 \setminus l_2|, |l_2 \setminus l_1| \leq t$ and $|l_1|, |l_2| \geq S - t$, where $l_1, l_2$ are as defined in the statement of Claim C. Combining all the above we get the lemma. ◇

We now proceed to prove the claims stated in the proof of Lemma B.

**Claim A** *We can have a coupling, call it $\mathcal{K}_2$, that chooses uniformly at random an element from each of the sets $\Omega'_C$ and $\Omega'_F$, such that the probability for the two chosen elements to specify different colour assignment for any vertex in $CH_u$ is upper bounded by $\sum_{w \in CH_u} p_w$.*

**Proof:** The coupling $\mathcal{K}_2$ is defined as follows: Choose u.a.r. a member of $\Omega_C$, let $C$ be the chosen element, by using $\mathcal{K}_1$, take the corresponding element of $\Omega_F$, let $C'$ be that element. Then, choose u.a.r. one among the copies of $C$ in $\Omega'_C$ and one of the copies of $C'$ in $\Omega'_F$. Clearly, each of the elements of $\Omega'_C$ and $\Omega'_F$ is chosen, uniformly at random. Our claim follows by noting that, the chosen elements of $\Omega'_C$ and $\Omega'_F$ differ in the colour assignments of the vertices in $CH_u$ iff $C$ and $C'$ do so. ◇

**Claim B** *Assume for the vertex $u$ that $|CH_u| = k$ and the disagreement probability for $w \in CH_u$ is $p_w$. If at the coupling of the second step of the processes ColourRoot in $\mathcal{C}(T, S, l)$, each $C \in \Omega'_C \cup \Omega'_F$ is assigned a weight $W_C$, then for the event $\mathcal{A}$ we have that*

$$Pr[\mathcal{A}] \leq \frac{1}{q_{k,S}} \frac{max_{C \in \Omega'_F \cup \Omega'_C}\{W_C\}}{min_{C \in \Omega'_F \cup \Omega'_C}\{W_C | W_C > 0\}} \sum_{w \in CH_u} p_w. \qquad (3)$$

*where $q_{k,S}$ is the probability of the event that after $k$ trials, not all elements of $[S]$ have been chosen, when at each trial we choose u.a.r. a member of $[S]$.*

**Proof:** Consider that we choose from $\Omega'_F$ such that the element $C$ is chosen with probability proportional to its weight, $W_C$. Consider the same for the set $\Omega'_C$. If there is a coupling of these two random weighted selections above, such that the probability of the event $\mathcal{A}$ to be upper bounded as in (3), then we are done.

The assumption that in the coupling $\mathcal{C}(T, \mathcal{S}, l)$, for each $w \in CH_u$, the disagreement probability is $p_w$, is equivalent to the following: There is a mapping, call it $f : \Omega'_F \to \Omega'_C$, which is one to one (and 'onto', since $|\Omega'_F| = |\Omega'_C|$) and for any pair of colourings $(C, f(C)) \in \Omega'_F \times \Omega'_C$ chosen u.a.r. the probability to specify different colourings for the vertices in $CH_u$ is upper bounded by $\sum_{w \in CH_u} p_w$.

Clearly the mapping $f$ defines a coupling for the "nonweighted" joint random selection of the elements the sets $\Omega'_F$ and $\Omega'_C$, since the two sets are equal sized. Based on $f$ we define a coupling for the "weighted" joint random selection of the elements of the sets $\Omega'_F$ and $\Omega'_C$.

From $\Omega'_i$ we produce the set $\Omega^W_i$, for $i \in \{C, F\}$, as follows: For each $C \in \Omega'_i$, insert into $\Omega^W_i$, $W_C$ copies of $C$, i.e. the elements $\{C_1, \ldots, C_{W_C}\}$. The weighted random selection from $\Omega'_i$ is equivalent to consider that we have chosen $C \in \Omega'_i$ if a random uniform selection from $\Omega^W_i$ have chosen one of $\{C_1, \ldots, C_{W_C}\}$. Thus, the construction of a coupling of the weighted joint selection from the sets $\Omega'_F$ and $\Omega'_C$ can, equivalently, be reduced to creating a coupling that selects uniformly at random one element from each of the sets $\Omega^W_F$ and $\Omega^W_C$. This is what are we doing in what follows.

First, we create a mapping $f' : (\Omega^W_F \cup \omega_2) \to (\Omega^W_C \cup \omega_1)$, where $\omega_2 \subset \Omega^W_F$ and $\omega_1 \subset \Omega^W_C$ and they will be defined soon after. The mapping $f'$ will be created based on the mapping $f$. Then we define the coupling which consists of choosing u.a.r. a member of $\Omega^W_F \cup \omega_2$ and then applying the chosen element to $f'$ so as to get a member of $\Omega^W_C \cup \omega_1$. In this coupling, the marginal probability for each member in $\Omega^W_F$ to be chosen, will be the same for all members. This should also hold for the members of $\Omega^W_C$. The claim will follow by bounding, appropriately, the quantity $Pr[\mathcal{A}]$.

We will define the sets $\omega_1, \omega_2$ as we construct $f'$. The mapping $f'$ is defined as follows: For each $C \in \Omega'_F$, with $f(C) = Q$ and $W_C = W_{f(C)} > 0$, set $f'(C_i) = Q_i$ for $i = 1, \ldots, W_C$. For each $C \in \Omega'_F$, with $W_C > W_{f(C)}$ and $f(C) = Q$, set $f'(C_i) = Q_i$ for $i = 1, \ldots, W_{f(C)}$ and for $i = W_{f(C)} + 1, \ldots, W_C$ set $f'(C_i)$ a u.a.r. chosen member of $\Omega^W_C$. Let $\omega_1$ be the set of all the elements of $\Omega^W_C$ that were randomly selected, as described above. For each $C \in \Omega'_F$, with $W_C < W_{f(C)}$ and $f(C) = Q$, we set $f'(C_i) = Q_i$ for $i = 1, \ldots, W_C$, and for $i = W_C + 1, \ldots, W_{f(C)}$, for the copy $Q_i$ choose u.a.r. a member of $\Omega'_F$ to correspond to. Let $\omega_2$ be the set of all the elements of $\Omega'_F$ that were randomly selected, as described above.

If we choose uniformly at random from $\Omega^W_F \cup \omega_2$, each element of $\Omega^W_F$ appears equiprobably. Similarly, if we choose u.a.r. from $\Omega^W_C \cup \omega_1$, each element of $\Omega^W_C$ appears equiprobably. Furthermore, if we choose u.a.r. from $\Omega^W_F \cup \omega_2$, and apply $f'$ to get a member from $\Omega^W_C \cup \omega_1$, all the members of $\Omega'_C \cup \omega_1$ have the same probability to be chosen, since the mapping $f'$ is one to one and onto ($|\Omega^W_F \cup \omega_2| = \Omega'_C \cup \omega_1$). Thus, in the coupling where we choose u.a.r. $\Omega^W_F \cup \omega_2$ and apply the mapping $f'$ and get a member of $\Omega'_C$, the marginal probability for all the members of $\Omega^W_F$ (and $\Omega^W_C$) to be chosen is the same.

What remains to be shown is that, in the coupling above, the event $\mathcal{A}$ occurs with probability $Pr[\mathcal{A}]$ which is upper bounded as in (3).

Clearly, for the colourings in the pairs $(C, f(C)) \in \Omega'_F \times \Omega'_C$ that define the same colour assignments for the vertices in $CH_u$, we have $W_C = W_{f(C)}$. For $C \in \Omega'_F$ in such a pair of colourings, it holds that the copy $C_i$, that $C$ has in $\Omega^W_F$, is corresponded through $f'$ to the copy $Q_i$, that $Q = f(C)$ has in $\Omega^W_C$, i.e. $Q_i = f(C_i)$, for $i = 1, \ldots, W_C$. Note that the event $\mathcal{A}$ does not hold for these pairs, $(C_i, f'(C_i))$ for $i = 1, \ldots, W_C$.

For each pair $(C, f(C)) \in \Omega'_F \times \Omega'_C$ that define a different colour assignments for the vertices in $CH_u$, it does not necessarily hold $W_C = W_{f(C)}$. Consider, first, the case where $W_C = W_{f(C)}$. Then, for $C \in \Omega'_F$ in such a pair of colourings, it holds that each copy $C_i$, that $C$ has in $\Omega^W_F$, is corresponded through $f'$ to the copy $Q_i$, that $Q = f(C)$ has in $\Omega^W_C$, for $i = 1, \ldots, W_C$. The event $\mathcal{A}$ does hold for these pairs, $(C_i, f'(C_i))$ for $i = 1, \ldots, W_C$.

Finally, we consider the case where the pair $(C, f(C)) \in \Omega'_F \times \Omega'_C$ defines a different colour assignments for the vertices in $CH_u$ and $W_C \neq W_{f(C)}$. W.l.o.g. we assume that $W_C > W_{f(C)}$. Then, for $C \in \Omega'_F$ in such a pair of colourings, it holds that each copy $C_i$, that $C$ has in $\Omega^W_F$, is corresponded through $f'$ to a copy of $Q_i$, that $Q = f(C)$ has in $\Omega^W_C$, for $i = 1, \ldots, W_{f(C)}$. The event $\mathcal{A}$ does hold for the pairs $(C_i, f'(C_i))$, $i = 1, \ldots, W_{f(C)}$. The remaining copies $C_i$, that $C$ has in $\Omega^W_F$, through $f'$ are mapped to u.a.r. chosen member of $\Omega^W_C$, for $i = W_{f(C)} + 1, \ldots, W_C$. Note that the event $\mathcal{A}$ does not necessarily hold for these pair. However, we assume that it does, which, clearly, is an overestimate for the probability $Pr[\mathcal{A}]$.

Let, $\Omega^W_A \subset \Omega^W_F \cup \omega_2$ be such that $\Omega^W_A = \{C \in \Omega^W_F \cup \omega_2 |$ for $(C, f'(C))$ the event $\mathcal{A}$ holds$\}$ and $\Omega_A \subset \Omega'_F$ be $\Omega_A = \{C \in \Omega'_F |$ for $(C, f(C))$ the event $\mathcal{A}$ holds$\}$. Clearly, $Pr[\mathcal{A}] = \frac{|\Omega^W_A|}{|\Omega^W_F \cup \omega_2|}$.

Let $\Omega_i^{(>0)} \subset \Omega'_i$ be such that $\Omega_i^{(>0)} = \{C \in \Omega'_i | W_C > 0\}$ and $q_i = \frac{|\Omega_i^{(>0)}|}{|\Omega'_i|}$, for $i \in \{C, F\}$. One can see that $|\Omega^W_A| \leq |\Omega_A| \max_{C \in \Omega'_C \cup \Omega'_F}\{W_C\}$, and $|\Omega^W_F \cup \omega_2| \geq |\Omega_F^{(>0)}| \cdot \min_{C \in \Omega_C^{>0} \cup \Omega_F^{>0}}\{W_C\}$. From the fact that $|\Omega_F^{\geq 0}| = q_F |\Omega_F|$ we get that

$$Pr[\mathcal{A}] \leq \frac{\max_{C \in \Omega'_C \cup \Omega'_F}\{W_C\}}{q_F \cdot \min_{C \in \Omega_C^{\geq 0} \cup \Omega_F^{\geq 0}}\{W_C\}} \frac{|\Omega_A|}{|\Omega_F|}.$$

Clearly, $q_F = q_{k,\mathcal{S}}$, where $q_{k,\mathcal{S}}$ is as defined in the statement of the claim. The claim follows by noting that $\frac{|\Omega_A|}{|\Omega_F|} \leq \sum_{w \in CH_u} p_w$. $\diamond$

**Claim C** *Consider the coupling $\mathcal{C}(T, \mathcal{S}, l)$ when it assigns colourings to the vertex $u$. Assume that the two processes chose members of $\Omega_F$ and $\Omega_C$ that specify colourings for the vertices in $CH_u$ such that for the vertex $u$, there are two lists of available colours, the $l_1$ and the $l_2$, correspondingly. Assuming that $|l_i| > 0$, for $i = 1, 2$, there is a coupling that can choose the same colour for the vertex $u$ with probability at least*

$$1 - \frac{max\{|l_1 \setminus l_2|, |l_2 \setminus l_1|\}}{min\{|l_1|, |l_2|\}}$$

.

**Proof:** The coupling that can choose the same colour for the vertex $u$ with probability indicated in the statement of the claim is the *maximal coupling* (see [10]).

More speciically, we assume, w.l.o.g., that $|l_1| \geq |l_2|$. Let $U$ be a random variable uniformly distributed in $(0, 1)$. We assume that if $\frac{i-1}{|l_1|} < U < \frac{i}{|l_1|}$ we choose the color $i \in l_1$, for $= 1, \ldots, |l_1|$. Also, $\forall i \in l_1 \cap l_2$ assume that if $\frac{i-1}{|l_1|} < U < \frac{i}{|l_1|}$ we choose $i$ in $l_2$. For $U$ everywhere else in $(0, 1)$ make an arbitrary arrangement so as each element of $l_2$ to be chosen with probability $1/l_2$. By the assumption that $|l_1| \geq |l_2|$, to all members $i \in l_1 \cap l_2$ we have assigned intervals which correspond to probability $1/|l_1| \leq 1/|l_2|$.

Clearly, the interval in $(0, 1)$ that corresponds to choosing different colourings from $l_1$ and $l_2$ is of length $\frac{|l_1 \setminus l_2|}{|l_1|}$. The claim follows by the fact that

$$\frac{|l_1 \setminus l_2|}{|l_1|} \leq \frac{max\{|l_1 \setminus l_2|, |l_2 \setminus l_1|\}}{min\{|l_1|, |l_2|\}}$$

**Lemma C** *Let $u$ be a vertex of the tree $T$ which is* nonmixing *and has $k$ children. If for every $w \in CH_u$ the probability of disagreement, in $\mathcal{C}(T, \mathcal{S}, l)$, is $p_w$, then for the vertex $u$, the probability of disagreement $p_u$, in $\mathcal{C}$, is bounded as*

$$p_u \leq S \frac{1}{q_{k,\mathcal{S}}} \left( \sum_{w \in CH_u} p_w \right) \qquad (4)$$

*and $q_{k,\mathcal{S}}$ is the probability of the event that after $k$ trials, not all elements of the set $[\mathcal{S}]$ have been chosen, when at each trial we choose uniformly at random a member of $[\mathcal{S}]$.*

**Proof:** Assume that in the coupling $\mathcal{C}(T, \mathcal{S}, l)$, when the vertex $u$ is to be coloured, the processes ColourRoot$(T, \mathcal{S}, C_1(V_1))$ and ColourRoot$(T, \mathcal{S})$ at their second step choose from the set of colourings $\Omega_C$ and $\Omega_F$, correspondingly.

Let $\mathcal{A}$ be the event that, when the vertex $u$ is to be coloured in $\mathcal{C}$ the members of $\Omega_F$ and $\Omega_C$, that are chosen at step (2) of the two processes ColourRoot, correspondingly, specify different colour assignments for the vertices in $CH_u$. Then

$Pr[\text{disagreement on } u] = Pr[\text{disagreement on } u | \mathcal{A}] Pr[\mathcal{A}] + Pr[\text{disagreement on } u | \overline{\mathcal{A}}] Pr[\overline{\mathcal{A}}]$

We mention that, if the event $\mathcal{A}$ does not hold ($\overline{\mathcal{A}}$ holds), then there is a coupling for step (3) of the ColourRoot that assigns the same colour to the vertex $u$ in $\mathcal{C}$, i.e. $Pr[\text{disagreement on } u | \overline{\mathcal{A}}] = 0$. Thus,

$$Pr[\text{disagreement on } u] = Pr[\text{disagreement on } u | \mathcal{A}] Pr[\mathcal{A}] \qquad (5)$$

To show the lemma we derive appropriate upper bounds for the probabilities in (5), $Pr[\mathcal{A}]$ and $Pr[\text{disagreement on } u | \mathcal{A}]$. We work exactly in the same manner as in the proof of Lemma B so as to get an upper bound for the term $Pr[\mathcal{A}]$, i.e. as in Lemma B we have

$$Pr[\mathcal{A}] \leq \frac{1}{q_{k,\mathcal{S}}} \frac{max_{C \in \Omega'_F \cup \Omega'_C} \{W_C\}}{min_{C \in \Omega'_F \cup \Omega'_C} \{W_C | W_C > 0\}} \sum_{w \in CH_u} p_w.$$

Note that at the coupling, of the step (2) of the ColourRoot, for colouring $u$, no member of either $\Omega'_C$ or $\Omega'_F$ is assigned weight more than $\mathcal{S}$ and the minimum non zero weight is 1. Furthermore, for a nonmixing vertex $u$, of sufficiently high degree, there are colourings of its children that use every colour in $[\mathcal{S}]$, these colourings are assigned weight zero, in in this case we have $q_{k,\mathcal{S}} \leq 1$.

The lemma follows by assuming that $Pr[\text{disagreement on } u | \mathcal{A}] = 1$ which is, clearly, an overestimate. ◇

## 3.2 The case of a tree - How to prove Lemma E.

Consider an instance of $G_{n,d/n}$, where $d > 1$, and for each vertex $v$ consider the graph $G_{v,d,\epsilon \log n}$, where $\epsilon = \frac{0.9}{\log(e^2 d/2)}$. By Lemma A it holds that w.h.p. $G_{v,d,\epsilon \log n}$ is either a unicyclic graph or a tree. Here, we condition that the graph $G_{v,d,\epsilon \log n}$ is a tree.

**Definition K** *The graph $G_{v,d,\epsilon \log n}$ when we condition that it is a tree, defines a probability space over the trees which we call $T_d$.*

Note that each nonleaf vertex of an instance of $T_d$ has a number of children whose distribution is dominated by $B(n, d/n)$, i.e. the binomial distribution with parameters $n$ and $d/n$.

Clearly, Lemma E will follow by showing that if $\mathcal{S}$ and $\epsilon$ are as large as specified by this lemma, then with probability, at least, $1 - 2n^{-1.25}$, for the tree $T$, an instance of $T_d$ rooted at the vertex $r$, the disagreement probability $p_r$ in the coupling $\mathcal{C}(T, \mathcal{S}, \lfloor \epsilon \log n \rfloor)$ is bounded as $p_r \leq n^{-1.25}/2$ (see Corollary C).

It is easy to see that if, $T$, an instance of $T_d$, is of height less than some integer $l$, then the disagreement probability on the root of $T$ in $\mathcal{C}(T, \mathcal{S}, l)$, is zero.

In $\mathcal{C}(T, \mathcal{S}, l)$, where $T$ is an instance of $T_d$ rooted at $r$, the disagreement probability $p_r$ depends only on the structure of the instance of $T_d$, for a given $\mathcal{S}$. We remind the reader that in $\mathcal{C}$, we assume that the boundary conditions are set so as to *maximize* the disagreement probability $p_r$. Clearly, $p_r$ is a *random variable*. We use the Lemma B and Lemma C to derive an upper bound for the expectation of $p_r$. The expectation of $p_r$ depends on $l$, $\mathcal{S}$ and $t$, the maximum number of children of a mixing vertex. Let $q(t)$ be the probability for a random variable, distributed as in $B(n-1, d/n)$, for fixed $d$, to be less than $t$.

**Lemma D** *For positive integers $\mathcal{S}$, $l$, real $d > 1$, in the coupling $\mathcal{C}(T, \mathcal{S}, l)$, where $T$ is an instance of $T_d$, the expectation of the disagreement probability $p_r$, on the root of $T$, is bounded as*

$$E[p_r] \leq \left( d \frac{t \cdot \mathcal{S}}{(\mathcal{S} - t)^2} q(t) + 2d \left( \mathcal{S}(1 - q(t)) + \frac{\mathcal{S}}{\mathcal{S} - 1} \left( \exp\left\{ \frac{d}{(\mathcal{S} - 1)} \right\} - q(t) \right) \right) \right)^l \qquad (6)$$

**Proof:** We remind the reader that $t$ stands for the maximum number of children of a mixing vertex. Let $q(t)$ be the probability for a random variable, distributed as in $B(n-1, d/n)$, for fixed $d$, to be less than $t$. Let

$$a(i) = \begin{cases} \dfrac{t \cdot \mathcal{S}}{(\mathcal{S} - t)^2} & \text{if } i \leq t \\ \dfrac{\mathcal{S}}{q_{i,\mathcal{S}}} & \text{otherwise} \end{cases}$$

where $q_{i,\mathcal{S}}$, is as defined in the statement of Lemma C.

Consider the coupling $\mathcal{C}(T, \mathcal{S}, l)$, where $T$ is an instance of $T_d$ rooted at the vertex $r$. Let $E[p_r]$ be the expectation of the disagreement probability on the root $r$. Conditioning on the number of children of $r$ and the disagreement probability $p_w$, $\forall w \in CH_r$ in $\mathcal{C}(T, \mathcal{S}, l)$, by Lemma B and Lemma C we have

$$E[p_r | p_w, \forall w \in CH_r] = a(|CH_r|) \sum_{w \in CH_r} p_w$$

By definition, $\forall w \in CH_r$, $p_w$ is upper bounded by the disagreement probability on the vertex $w$ in the coupling $\mathcal{C}(T_w, \mathcal{S}, l-1)$ where $T_w$ is the subtree of $T$ rooted at vertex $w$. Call this disagreement probability $p_w^*$. We clear out that $p_w$ refers to the coupling $\mathcal{C}(T, \mathcal{S}, l)$ while $p_w^*$ to $\mathcal{C}(T_w, \mathcal{S}, l-1)$. It is direct that

$$E[p_r] \leq \sum_{i=0}^{n} i a(i) Pr[|CH_r| = i] E[p_w^*] \qquad \text{for } w \in CH_r$$

Note that the random variables $p_w^*$ for $w \in CH_r$ are identically distributed and independent of the number of children of $r$. Also, noting that the function $f(i) = i \cdot a(i)$ is increasing for $t \ll \mathcal{S}$ and by the fact that the distribution of the number of children of $r$ is dominated by the $B(n, d/n)$, (by proposition 9.1.2. of [13]), it holds that

$$E[p_r] \leq \sum_{i=0}^{n} i \cdot a(i) \binom{n}{i} p^i (1-p)^{n-i} E[p_w^*] \qquad \text{for } w \in CH_r \tag{7}$$

where $p = d/n$. Let $S_1 = \sum_{i=0}^{t} i \cdot a(i) \binom{n}{i} p^i (1-p)^{n-i}$ and $S_2 = \sum_{i=t+1}^{n} i \cdot a(i) \binom{n}{i} p^i (1-p)^{n-i}$.

$$\begin{aligned} S_1 &\leq \frac{t \cdot \mathcal{S}}{(\mathcal{S}-t)^2} \sum_{i=0}^{t} i \binom{n}{i} p^i (1-p)^{n-i} \\ &= \frac{t \cdot \mathcal{S}}{(\mathcal{S}-t)^2} np \sum_{i=0}^{t-1} \binom{n-1}{i} p^i (1-p)^{n-1-i} \\ &= \frac{t \cdot \mathcal{S}}{(\mathcal{S}-t)^2} q(t) d \end{aligned}$$

Before calculating $S_2$, we eliminate the probability term $q_{i,\mathcal{S}}$ from the $a(i)$ for $i > t$. For $q_{i,\mathcal{S}}$ it holds that

$$q_{i,\mathcal{S}} \geq \mathcal{S}\left(1 - \frac{1}{\mathcal{S}}\right)^i \left(1 - q_{i,(\mathcal{S}-1)}\right)$$

i.e. the probability of the event "not choosing *some* element of $[\mathcal{S}]$ after $i$ trials" is greater than, or equal to the probability of the event "not choosing *exactly* one element of $[\mathcal{S}]$", since the second event is a special case of the first one. Furthermore, since $q_{k,(\mathcal{S}-1)} \leq q_{k,\mathcal{S}}$ we get that

$$q_{i,\mathcal{S}} \geq \mathcal{S}\left(1 - \frac{1}{\mathcal{S}}\right)^i (1 - q_{i,\mathcal{S}})$$

Let $\Omega = \{1, \ldots, n\}$ and let $t_0 = \sup\{t \in \Omega \mid q_{t,\mathcal{S}} \geq 1/2\}$. Instead of using $q_{i,\mathcal{S}}$ we make the following simplification. For $i > t_0$ we bound the quantity $1/q_{i,\mathcal{S}}$ as

$$\frac{1}{q_{i,\mathcal{S}}} \leq \frac{2}{\mathcal{S}\left(1-\frac{1}{\mathcal{S}}\right)^i} = \frac{2}{\mathcal{S}}\left(\frac{\mathcal{S}}{\mathcal{S}-1}\right)^i.$$

Also, for $i \leq t_0$, clearly, $1/q_{i,\mathcal{S}} \leq 2$.

$$\begin{aligned} S_2 &\leq 2\mathcal{S} \sum_{i=t+1}^{t_0} i \binom{n}{i} p^i (1-p)^{n-i} + 2 \sum_{i=t_0+1}^{n} i \binom{n}{i} \left(\frac{\mathcal{S}}{\mathcal{S}-1}\right)^i p^i (1-p)^{n-i} \\ &\leq 2\mathcal{S} \sum_{i=t+1}^{n} i \binom{n}{i} p^i (1-p)^{n-i} + 2 \sum_{i=t+1}^{n} i \binom{n}{i} \left(\frac{\mathcal{S}}{\mathcal{S}-1}\right)^i p^i (1-p)^{n-i} \\ &\leq 2\mathcal{S} np \sum_{i=t}^{n-1} \binom{n-1}{i} p^i (1-p)^{n-1-i} + 2np \frac{\mathcal{S}}{\mathcal{S}-1} \sum_{i=t}^{n-1} \binom{n-1}{i} \left(\frac{\mathcal{S}}{\mathcal{S}-1}\right)^i p^i (1-p)^{n-1-i} \\ &\leq 2\mathcal{S} d \left(1 - \sum_{i=0}^{t-1} \binom{n-1}{i} p^i (1-p)^{n-1-i}\right) + \\ &\quad + 2\frac{\mathcal{S}}{\mathcal{S}-1} d \left(\left(1 - p + \frac{\mathcal{S}}{\mathcal{S}-1} p\right)^{n-1} - \sum_{i=0}^{t-1} \binom{n-1}{i} \left(\frac{\mathcal{S}}{\mathcal{S}-1}\right)^i p^i (1-p)^{n-1-i}\right) \\ &\leq 2\mathcal{S} d (1 - q(t)) + 2\frac{\mathcal{S}}{\mathcal{S}-1} d \left(\left(1 + \frac{1}{\mathcal{S}-1} p\right)^{n-1} - \sum_{i=0}^{t-1} \binom{n-1}{i} p^i (1-p)^{n-1-i}\right) \\ &\leq 2\mathcal{S} d (1 - q(t)) + 2\frac{\mathcal{S}}{\mathcal{S}-1} d \left(\left(1 + \frac{1}{\mathcal{S}-1} p\right)^{n-1} - q(t)\right) \\ &\leq 2d \left(\mathcal{S}(1 - q(t)) + \frac{\mathcal{S}}{\mathcal{S}-1} (\exp\{d/(\mathcal{S}-1)\} - q(t))\right) \end{aligned}$$

Substituting the bounds for $S_1$ and $S_2$ in (7) we get

$$E[p_r] \leq \left(d\frac{t\mathcal{S}}{(\mathcal{S}-t)^2}q(t) + 2d\left(\mathcal{S}(1-q(t)) + \frac{\mathcal{S}}{\mathcal{S}-1}\left(\exp\{d/(\mathcal{S}-1)\} - q(t)\right)\right)\right) E[p_w^*]$$

for $w \in CH_r$. We can substitute $E[p_w^*]$ in the same manner as $E[p_r]$. Using induction and assuming that for the vertices at distance $l$ from the root the expectation of the probability of disagreement is 1, the lemma follows. $\diamond$

Finally, Lemma E follows by setting appropriate quantities for $\mathcal{S}$ and $l$, in (6) and then applying the Markov inequality. Here it is crucial to remark that if $d$ is sufficiently large, then for $t \geq 7d$ it holds $q(t) \geq 1 - d^{-28}$.

**Lemma E** *Consider a system $PCS(G_{v,d,\epsilon \log n}, \mathcal{S})$, for $d > 1$, $\epsilon = \frac{0.9}{4\log(e^2 d/2)}$ and for $G_{v,d,\epsilon \log n}$ we assume that it is a tree. If the cardinality of $\mathcal{S}$ is a sufficiently large constant, then with probability at least $1 - 2n^{-1.25}$, for the above system it holds that $SD(v, \lfloor \epsilon \log n \rfloor) = n^{-1.25}$. For sufficiently large $d$, we should have $\mathcal{S} \geq d^{14}$.*

**Proof:** By Lemma D we have that, in the coupling $\mathcal{C}(G_{v,d,\epsilon \log n}, \mathcal{S}, C(L))$, the expectation of $p_v$, is bounded as

$$E[p_v] \leq \left(d\frac{t \cdot \mathcal{S}}{(\mathcal{S}-t)^2}q(t) + 2d\left(\mathcal{S}(1-q(t)) + \frac{\mathcal{S}}{\mathcal{S}-1}\left(\exp\{d/(\mathcal{S}-1)\} - q(t)\right)\right)\right)^l \qquad (8)$$

where $l$ is the minimum distance of $v$ and the boundary set $L$, $q(t)$ is equal to the probability for a random variable, distributed as in $B(n-1, d/n)$, for fixed $d$, to be less than $t$, the maximum number of children of a mixing vertex.

Set $l = \epsilon \log n$, with $\epsilon = \frac{0.9}{\log(e^2 d/2)}$ in (8). To prove the lemma it suffices to show that for $\mathcal{S}$ as described in the statement of the lemma and appropriately large $t$ we have $E[p_v] \leq n^{-2.5}$. Clearly, for $E[p_v] \leq n^{-2.5}$ by using the Markov Inequality (see [3]) we can get that

$$Pr[p_v \geq 2n^{-1.25}] \leq \frac{E[p_v]}{2n^{-1.25}} \leq n^{-1.25}/2$$

By Definition F and Theorem D we get that if $E[p_v] \leq n^{-2.5}$, then with probability at least $1 - Pr[p_v \geq 2n^{-1.25}] \geq 1 - 2n^{-1.25}$ for the system $PCS(G_{v,d,\epsilon \log n}, \mathcal{S})$ it holds that $SD(v, \lfloor \epsilon \log n \rfloor) \leq n^{-1.25}$, which proves the lemma.

Thus, what remains to be shown is that there are appropriate values for $t$ and $\mathcal{S}$ such that, $E[p_v] \leq n^{-2.5}$.

First, we show that if $d$ is a sufficiently large constant, then for $\mathcal{S} \geq d^{14}$ and $t$ such that $q(t) \geq 1 - d^{-28}$ we get $E[p_v] \leq n^{-2.5}$. Using Corollary B (from [6]) we see that when $t \geq \max\{7d, 28\log d + 1\}$ it holds $q(t) \geq 1 - d^{-28}$.

**Corollary B** *If a random variable $Z$ is distributed as in $B(n,q)$ with $\lambda = nq$ then*

$$Pr[Z \geq x] \leq e^{-x} \quad x \geq 7\lambda. \qquad (9)$$

For a proof of the above corollary, see in [6] the proof of Corollary 2.4.

Assuming that $d$ is a sufficiently large constant, we substitute the values of $\mathcal{S}$ with $d^{14}$ and $t = 7d$, in (8) and we get

$$E[p_v] \leq \left(\frac{7d^{16}}{(d^{14}-7d)^2} + 2d\left(d^{14}d^{-28} + \frac{d^{14}}{d^{14}-1}\left(1 + \frac{d}{d^{14}-1} + \frac{e^\xi}{2!}\frac{d^2}{(d^{14}-1)^2} - 1 + d^{-28}\right)\right)\right)^{\epsilon \log n}$$

where $0 < \xi < d/(d^{14} - 1)$. In the above inequality we used the fact that $1 - d^{-28} \leq q(t) \leq 1$, and substituted $e^{d/(S-1)}$ by using the MacLaurin series of the function $f(x) = e^x$, for real $x$. Thus, we get

$$E[p_v] \leq \left( \frac{7d^{-12}}{(1 - 7d^{-13})^2} + 2d \left( d^{-14} + \frac{1}{1 - d^{-14}} \left( \frac{d^{-13}}{1 - d^{-14}} + \frac{e}{2} \frac{d^{-26}}{(1 - d^{-14})^2} + d^{-28} \right) \right) \right)^{\epsilon \log n}$$

$$\leq \left( d^{-12} \left( \frac{7}{(1 - 7d^{-13})^2} + 2d^{-1} + \frac{2}{(1 - d^{-14})^2} + \frac{ed^{-13}}{(1 - d^{-14})^3} + \frac{2d^{-15}}{1 - d^{-14}} \right) \right)^{\epsilon \log n}$$

Taking $d$ at least 20, we get that

$$E[p_v] \leq n^{\epsilon \log(9.2 d^{-12})}$$

Replacing $\epsilon$, we see that it suffices to hold $0.9 \log(9.2 d^{-12}) \leq -10 \log(e^2 d/2)$, or $9.2 d^{-12} \leq (e^2 d/2)^{-11.11}$ which clearly holds for sufficiently large constant $d$.

For relatively smaller $d$, one can easily see that setting $S = d^x$, for an appropriately large constant exponent $x$ and arranging the quantity $t$ so as $q(t) \geq 1 - d^{-2x}$ similarly to the previous case, can get

$$E[p_v] \leq \left( d^{-x+2} \left( \frac{t/d}{(1 - d^{-x} t)^2} + 2d^{-1} + \frac{2d^{-x-1}}{1 - d^{-x}} + \frac{2}{(1 - d^{-x})^2} + e \frac{d^{-x+1}}{(1 - d^{-x})^3} \right) \right)^{\epsilon \log n}$$

We take $x$ sufficiently large so as to have $1 - d^{-x} \geq 1 - 10^{-3}$ and $xd^{-x} \leq 10^{-3}$.

If $t = 7d$, then, with the above assumptions, we can easily derive that $E[p_v] \leq (d^{-x+2} 16)^{\epsilon \log n}$. For this case, if $E[p_v] \leq n^{-2.5}$, then we should have $16 d^{-x+2} \leq (e^2 d/2)^{11.11}$, which clearly holds for sufficiently large $x$.

If $2x \log d + 1 > 7d$, then by Corollary B we should have $t = 2x \log d + 1$. With the assumptions we have made for $x$ we get that $E[p_v] \leq (d^{-x+2}(2.1 x \frac{\log d}{d} + 9))^{\epsilon \log n}$. If $E[p_v] \leq n^{-2.5}$, then it should hold $(d^{-x+2}(2.1 x \frac{\log d}{d} + 9)) \leq (e^2 d/2)^{11.11}$, which clearly holds for sufficiently large $x$.

The lemma follows

$\diamond$

### 3.3 The case of a unicyclic graph - How to prove Lemma H.

Consider an instance of $G_{n,d/n}$, and the set of its subgraphs $G_{v,d,\epsilon \log n}$, as in 3.2. By Lemma A, we have that w.h.p. $G_{v,d,\epsilon \log n}$ is either a unicyclic graph or a tree. Here, we condition that that $G_{v,d,\epsilon \log n}$ is a unicyclic graph.

First, we show how can we extend the techniques for proving Lemma E, i.e. proving spatial mixing properties of system with underlying graph a tree, to showing Lemma H, which refers to systems with a unicyclic underlying graph.

Consider the depth first search in $G_{v,d,\epsilon \log n}$, that starts from the vertex $v$ and let $u$ be the first vertex of the unique cycle that is reached by the search. Clearly, there are two possible choices for this search to explore the vertices of the cycle that $u$ belongs to. If $w_1$ and $w_2$ are the vertices on the cycle that are also adjacent to $u$, then let $T^1$ and $T^2$ be the two depth-first search trees of $G_{v,d,\epsilon \log n}$, rooted at $v$, with the first tree having $u$ adjacent only to $w_1$ (not adjacent to $w_2$) and the second having $u$ adjacent only to $w_2$.

**Definition L** *With the above notation, the tree $T_{r,d,\epsilon \log n}$ is isomorphic to the tree that comes up from the union of $T^1$ and $T^2_{w_2}$ plus an edge connecting the vertices $u$ in $T^1$ with $w_2$ in $T^2_{w_2}$. The root $r$ of $T_{r,d,\epsilon \log n}$ corresponds to the vertex $v$ in $T^1$.*

Note that the number of children of a nonleaf vertex of $T_{r,d,\epsilon \log n}$ has distribution which is dominated by the $B(n, d/n)$ with the condition that it is at least 2.

Each of the trees $T^1$ and $T^2_{w_2}$, in the definition of $T_{r,d,\epsilon \log n}$, are isomorphic to some subgraph of $G_{v,d,\epsilon \log n}$, i.e. there is a correspondence between the vertices in $T^1$ and $T^2_{w_2}$ with the vertices in $G_{v,d,\epsilon \log n}$. Based on this correspondence, we can define a function $h : V_T \to V_G$ where, $V_T$ is the set of vertices of $T_{r,d,\epsilon \log n}$ and $V_G$ the set of vertices in $G_{v,d,\epsilon \log n}$.

Let $L$ be the set that contains all the vertices in $G_{v,d,\epsilon \log n}$ that are at graph distance, at least, $\lfloor \epsilon \log n \rfloor$ from $v$. Consider the $\mathcal{S}$-colouring $C_1(L)$ which is such that, the set the total variation distance of the Gibbs measures $\mu(X_v|C_1(L))$ and $\mu(X_v)$, as these are specified by the system $PCS(G_{v,d,\epsilon \log n}, \mathcal{S})$, is maximized.

For the tree $T_{r,d,\epsilon \log n}$ derived by $G_{v,d,\epsilon \log n}$, the integers $\mathcal{S}$ and $l$, let $\mathcal{C}'(T_{r,d,\epsilon \log n}, \mathcal{S}, l)$, be the coupling of the stochastic processes ColourRoot$(T_{r,d,\epsilon \log n}, \mathcal{S}, C_1^T(A))$ and ColourRoot$(T_{r,d,\epsilon \log n}, \mathcal{S})$. The set $A$ is such that $\forall \hat{v} \in A \ \exists \hat{u} \in (V_1 \cup L)$ such that $h(\hat{v}) = \hat{u}$ and $C_1^T(\hat{v}) = C_1(\hat{u})$. The difference of $\mathcal{C}'$ from $\mathcal{C}$ regarding the coupling of the processes ColourRoot is that, if for a non mixing vertex $u$ in $T$, which has $i$ children, each vertex $w \in CH_u$ has disagreement probability $p_w$, then the disagreement probability $p_u$, is bounded as

$$p_u \leq \frac{\mathcal{S}}{q_{i,\mathcal{S},2}} \sum_{w \in CH_u} p_w. \tag{10}$$

where $q_{i,\mathcal{S},2}$ is the probability of the event that after $k$ trials, not all elements of the set $[\mathcal{S}]$ have been chosen, when at each trial we choose u.a.r. a member of $[\mathcal{S}]$ and conditioning that the first two trials chose different elements of $[\mathcal{S}]$.

Comparing the bound in (10) with that was given in (4) in the statement of Lemma C, it is easy to see that $q_{i,\mathcal{S},2} \leq q_{i,\mathcal{S}}$. This implies that the coupling $\mathcal{C}'$ can exist as, for all the vertices of $T$, it gives the same, or worse bounds for the probabilities of disagreement than $\mathcal{C}$, on the same input.

**Lemma F** *Consider the graph $G_{v,d,\epsilon \log n}$ and the corresponding tree $T_{r,d,\epsilon \log n}$, for $d > 1$ and $\epsilon = \frac{0.9}{4 \log(e^2 d/2)}$. If $p_r$ is the bound for the disagreement probability that we derive from Lemma B and (10) for the coupling $\mathcal{C}'(T_{r,d,\epsilon \log n}, \mathcal{S}, \lfloor \epsilon \log n \rfloor)$, then it holds that $SD(v, \lfloor \epsilon \log n \rfloor) \leq 2p_r$ in the system $PCS(G_{v,d,\epsilon \log n}, \mathcal{S})$, for any positive integer $\mathcal{S}$.*

**Proof:** Let $u$ be the vertex in $G_{v,d,\epsilon \log n}$ which belongs to the unique cycle of $G_{v,d,\epsilon \log n}$ and among all the vertices in the cycle it has the smallest distance from $v$. Let $G^u$ be the *connected* subgraph of $G_{v,d,\epsilon \log n}$ that contains the vertex $u$ and the vertices whose distance from $v$ is greater than that of the vertex $u$ from $v$. It is easy to see that that $G^u$ is a unicyclic graph and $G_{v,d,\epsilon \log n} \backslash G^u$ is a tree.

Let $L$ be the set that contains all the vertices in $G_{v,d,\epsilon \log n}$ that are at graph distance, at least, $\lfloor \epsilon \log n \rfloor$ from $v$. Consider the $\mathcal{S}$-colouring $C_1(L)$ which is such that the total variation distance of the Gibbs measures $\mu(X_v|C_1(L))$ and $\mu(X_v)$, as these are specified by the system $PCS(G_{v,d,\epsilon \log n}, \mathcal{S})$, to be maximized.

Assume that there is a coupling such that choosing uniformly at random a member from each of the sets of $\mathcal{S}$-colourings $\Omega(G^u, \mathcal{S}, C_1(L))$ and $\Omega(G^u, \mathcal{S})$, correspondingly, the probability for the two members to specify a different colour assignment for the vertex $u$ is $Q$. Clearly, using Lemma B and (10) we are able to derive an upper bound for the disagreement probability $p_v$, of the coupling of the processes ColourRoot$(T, \mathcal{S}, C_1(T \cap L))$ and ColourRoot$(T, \mathcal{S})$, where $T = G_{v,d,\epsilon \log n} \backslash (G^u \backslash \{u\})$ and with the assumption that the disagreement probability of the vertex $u$ is set apriori to $Q$. Note that the graph $G_{v,d,\epsilon \log n} \backslash (G^u \backslash \{u\})$ is a tree, i.e. it includes all the vertices in $G_{v,d,\epsilon \log n}$ apart from the vertice in $G^u \backslash \{u\}$. This bound of $p_v$ is also an upper bound for the total variation distance of the Gibbs measures, $\mu(X_v | C_1(L))$ and $\mu(X_v)$, as these are specified by the system $PCS(G_{v,d,\epsilon \log n}, \mathcal{S})$.

The lemma will follow by showing that for an appropriately constructed tree $T^u$, with respect to $G^u$, and appropriate boundary condition $C_1'$, if in coupling of the processes ColourRoot$(T^u, \mathcal{S}, C_1')$ and ColourRoot$(T^u, \mathcal{S})$ has disagreement probability $P$ at the vertex $u$, the root of $T^u$, then it holds $Q \leq P$, where the probability $Q$ is as previously defined. W.l.o.g. assume that the vertex $v$ belongs to the unique cycle of $G_{v,d,\epsilon \log n}$, i.e. the $G^u$ and $G_{v,d,\epsilon \log n}$ are identical.

Each of the trees $T^1$ and $T^2_{w_2}$, in the definition of $T_{r,d,\epsilon \log n}$ (Definition L), are isomorphic to some subgraph of $G_{v,d,\epsilon \log n}$, i.e. there is a correspondence between the vertices in $T^1$ and $T^2_{w_2}$ with the vertices in $G_{v,d,\epsilon \log n}$. Based on this correspondence, we can have a function $h : V_T \to V_G$ where, $V_T$ is the set of vertices of $T_{r,d,\epsilon \log n}$ and $V_G$ the set of vertices in $G_{v,d,\epsilon \log n}$.

Let $L'$ be the set of vertices in $T_{r,d,\epsilon \log n}$ such that $L' = \{u \in V_T | h(u) \in L\}$. It is direct that the vertex set $L'$ is at distance, at least, $\lfloor \epsilon \log n \rfloor$ from $r$, in $T_{r,d,\epsilon \log n}$. If $N_v$ is the vertex set that contains all the adjacent vertices of $v$ in $G_{v,d,\epsilon \log n}$, then for $w_0 \in N_v$, let $G^{w_0}$ be the *connected* component of $G_{v,d,\epsilon \log n} \backslash \{v\}$ that contains $w_0$. It is straightforward that the component $G^{w_0}$ is a tree which is isomorphic to the subtree $T_w$ of $T_{r,d,\epsilon \log n}$, where $h(w) = w_0$.

Let $C_T^i(L')$ be a colouring such that, for each vertex $w \in L'$ it specifies the same colour assignment as $C_i(L)$ specifies for $w_0 \in L$, where $h(w) = w_0$, for $i = 1, 2$. The fact, that, for any $w_0 \in N_v$ and $w \in CH_r$ such that $h(w) = w_0$, the graphs $G^{w_0}$ and $T_w$ are isomorphic implies that there is a correspondence between the elements of the sets $\Omega(G^{w_0}, \mathcal{S}, C_1(L))$ and $\Omega(T_w, \mathcal{S}, C_T^1(L'))$ such that for any two corresponding colourings $C^1$ and $C^2$ it holds that $\forall u_1 \in T_w \ C^1(h(u_1)) = C^2(u_1)$. Clearly, there is a similar correspondence between the members of the sets of $\mathcal{S}$-colourings $\Omega(G^{w_0}, \mathcal{S})$ and $\Omega(T_w, \mathcal{S})$, for $w_0 \in N_v$, $w \in CH_r$ and $h(w) = w_0$.

Using the above correspondences between the pairs of sets $\Omega(G^{w_0}, \mathcal{S}, C_1(L))$, $\Omega(T_w, \mathcal{S}, C_T^1(L'))$ and $\Omega(G^{w_0}, \mathcal{S})$, $\Omega(T_w, \mathcal{S}, C_T^2(L'))$, with the assumption that $\forall w \in CH_r$, in the coupling $\mathcal{C}'$ $(T_{r,d,\epsilon \log n}, \mathcal{S}, \lfloor \epsilon \log n \rfloor)$ the disagreement probability for the vertex $w$ is upper bounded by $p_w$, we get the following: There is a coupling such that choosing u.a.r. from the sets of $\mathcal{S}$-colourings $\Omega(G^{w_0}, \mathcal{S}, C_1(L))$ and $\Omega(G^{w_0}, \mathcal{S})$ such that the probability for the two chosen elements to assign different colour to the vertex $w_0$ is upper bounded by $p_w$, where $h(w) = w_0$. Furthermore, there is a coupling that chooses u.a.r. from the sets $\Omega(G_{v,d,\epsilon \log n} \backslash \{v\}, \mathcal{S}, C_1(L))$ and $\Omega(G_{v,d,\epsilon \log n} \backslash \{v\})$ and there is at least one vertex $N_v$ that the two choices specify a different colour assignment with probability upper bounded as $\sum_{w \in CH_r} p_w$. The bound for this probability is derived by using the union bound.

**Remark.** Note that if $w_1$ and $w_2$ are the vertices in $N_v$ that also belong to the cycle, then the events $e_i$ *"there is disagreement on the vertex $w_i$"*, for $i = 1, 2$, are correlated. However, the union bound we used for bounding the probability for at least one vertex in $N_v$ to be differently coloured, in the coupling, still holds even for correlated events.

**Claim D** *If, the coupling $\mathcal{C}'(T_{r,d,\epsilon \log n}, \mathcal{S}, \lfloor \epsilon \log n \rfloor)$, has probability disagreement on the vertex $r$ upper bounded by $p_r$, then we can have a coupling that chooses u.a.r. a member from each of the two sets $\Omega(G_{v,d,\epsilon \log n}, \mathcal{S}, C_1(L))$ and $\Omega(G_{v,d,\epsilon \log n}, \mathcal{S})$ such that, the probability for $v$ to have different colour assignments to be upper bounded by $p_r$.*

With the above claim and what follows, we get the proof of the lemma. Let $\tilde{C}(L)$ and $\hat{C}(L)$ be the two colourings which maximize the total variation distance of the measures $\mu(X_v | C(\tilde{C}(L))$ and $\mu(X_v|\hat{C}(L))$, as these are specified by the system $PCS(G_{v,d,\epsilon \log n}, \mathcal{S})$.

$$\begin{aligned} SD(v, \epsilon \log n) &= d_{TV}\left(\mu(X_v|C(\tilde{C}(L)), \mu(X_v|\hat{C}(L))\right) \\ &\leq d_{TV}\left(\mu(X_v|C(\tilde{C}(L)), \mu(X_v)\right) + d_{TV}\left(\mu(X_v), \mu(X_v|\hat{C}(L))\right) \\ &\leq 2 p_r \end{aligned}$$

where $p_r$ is the bound of the disagreement probability on the vertex $r$, at the coupling $\mathcal{C}'(T_{r,d,\epsilon \log n}, \mathcal{S}, \lfloor \epsilon \log n \rfloor)$. ◇

We now prove the claim that appears in the proof of Lemma F.

**Claim D** *If, the coupling $\mathcal{C}'(T_{r,d,\epsilon \log n}, \mathcal{S}, \lfloor \epsilon \log n \rfloor)$, has probability disagreement on the vertex $r$ upper bounded by $p_r$, then we can have a coupling that chooses u.a.r. a member from each of the two sets $\Omega(G_{v,d,\epsilon \log n}, \mathcal{S}, C_1(L))$ $\Omega(G_{v,d,\epsilon \log n}, \mathcal{S})$ such that, the probability for $v$ to have different colour assignments to be upper bounded by $p_r$.*

**Proof:** Assume that in the coupling $\mathcal{C}'(T_{r,d,\epsilon \log n}, \mathcal{S}, \lfloor \epsilon \log n \rfloor)$ each vertex of $T_{r,d,\epsilon \log n}$, whose number of children is at most $t'$, for some positive integer $t'$, is considered mixing. The vertices $r$ and $v$ have the same degree in $G_{v,d,\epsilon \log n}$ and $T_{r,d,\epsilon \log n}$, correspondingly.

Assuming that $\forall w \in CH_r$, in the coupling $\mathcal{C}'(T_{r,d,\epsilon \log n}, \mathcal{S}, \lfloor \epsilon \log n \rfloor)$ the disagreement probability for the vertex $w$ is upper bounded by $p_w$, we can see that there is a coupling that chooses u.a.r. a member from each of the sets $\Omega(G_{v,d,\epsilon \log n} \backslash \{v\}, \mathcal{S}, C_1(L))$ and $\Omega(G_{v,d,\epsilon \log n} \backslash \{v\}, \mathcal{S})$ and there is at least one vertex $N_v$ that the two choices specify different colour assignments with probability upper bounded as $\sum_{w \in CH_r} p_w$. Thus, the claim follows if, based on the above assumption, we give a coupling that selects u.a.r. a member from $\Omega(G_{v,d,\epsilon \log n}, \mathcal{S}, C_1(L))$ and $\Omega(G_{v,d,\epsilon \log n}, \mathcal{S})$ such that, the probability for $v$ to have different colour assignments to be upper bounded by the probability of disagreement $p_r$ in the coupling $\mathcal{C}'(T_{r,d,\epsilon \log n}, \mathcal{S}, \epsilon \log n)$.

If the degree of the vertex $v$ is at most $t'$, then we use Lemma B to get an upper bound for the probability for a coupling that chooses u.a.r. from the sets $\Omega(G_{v,d,\epsilon \log n}, \mathcal{S}, C_1(L))$ $\Omega(G_{v,d,\epsilon \log n}, \mathcal{S})$ to choose two members that specify different colour assignments for the vertex $v$. Clearly, we get the same bound for the probability of disagreement $p_r$ in the coupling $\mathcal{C}'(T_{r,d,\epsilon \log n}, \mathcal{S}, \lfloor \epsilon \log n \rfloor)$.

We note that despite the fact that the colour assignments of two adjacent vertices of $v$ are correlated, we can still apply Lemma B. This is because in Lemma B it is assumed that, if there is a disagreement in the vertices in $CH_v$ (to be exact with the context of the proof we have to write $N_v$), then all the vertices have different colour assignments. This leads clearly to an overestimate for bounding the disagreement probability for $p_v$ even for the case where the two colourings are correlated.

If the degree of the vertex $v$ is $i$, greater than $t'$, then we use Lemma C, with a little modification, to get an upper bound for the probability for a coupling that chooses u.a.r. from the sets $\Omega(G_{v,d,\epsilon \log n}, \mathcal{S}, C_1(L))$ and $\Omega(G_{v,d,\epsilon \log n}, \mathcal{S})$ to choose two members that specify different colour assignments for the vertex $v$. One can see that the term $1/q_{i,\mathcal{S}}$ in (4) of the statement of Lemma C is not exact for our case. More specifically, in the last paragraph of the proof of Claim B, for our case the quantity $q_F$ is not equal to $q_{i,\mathcal{S}}$ due to the fact that two vertices do not choose independently their colour assignment. However, it is direct that $q_F$ is lower bounded by the probability of the event that after $i$ trials, not all the elements of $[\mathcal{S}]$ have been chosen, when at each trial we choose u.a.r. a member of $[\mathcal{S}]$ and conditioning that the first two trials choose different elements of $[\mathcal{S}]$. With this modification we can see that the coupling of choosing u.a.r. from $\Omega(G_{v,d,\epsilon \log n}, \mathcal{S}, C_1(L))$ and $\Omega(G_{v,d,\epsilon \log n}, \mathcal{S})$ can be done such that the vertex $v$ to have different colour assignments with probability

$$p_v \leq \frac{\mathcal{S}}{q_{i,\mathcal{S},2}} \sum_{w \in CH_r} p_w.$$

where $q_{i,\mathcal{S},2}$ is the probability of the event that after $k$ trials, not all elements of $[\mathcal{S}]$ have been chosen, when at each trial we choose u.a.r. a member of $[\mathcal{S}]$ and conditioning that the first two trials chose different elements of $[\mathcal{S}]$.

The claim follows. ◇

Towards proving Lemma H, we use Lemma F which allows us to consider the tree $T_{r,d,\epsilon \log n}$ derived by unicyclic graph $G_{v,d,\epsilon \log n}$, instead of $G_{v,d,\epsilon \log n}$. In turn, we can use the same techniques as in section 3.2 for showing the desired spatial mixing properties for systems with underlying graph the tree $T_{r,d,\epsilon \log n}$. Note that now the coupling is $\mathcal{C}'$. Let $q(t)$ be equal to the probability for a random variable, distributed as in $B(n-1, d/n)$, for fixed $d$, to be less than $t$.

**Lemma G** *For positive integers $\mathcal{S}$, $l$, real $d > 1$, in the coupling $\mathcal{C}'(T, \mathcal{S}, l)$, where $T$ is an instance of $T_{r,d,\epsilon \log n}$, the expectation of the disagreement probability $p_r$, on the root of $T$, is bounded as*

$$E[p_r] \leq \left( \frac{1}{1 - (d+1)e^{-d}} \left( d \frac{t \cdot \mathcal{S}}{(\mathcal{S}-t)^2} q(t) + 2d \left( \mathcal{S}(1 - q(t)) + \exp\left\{\frac{d}{\mathcal{S}-2}\right\} - q(t) \right) \right) \right)^l$$

**Proof:** We remind the reader that $t$ stands for the maximum number of a children of a mixing vertex. Let $q(t)$ be the probability for a random variable, distributed as in $B(n-1, d/n)$, for fixed $d$, to be less than $t$.

One can see than in the $T_{r,d,\epsilon \log n}$, the number of children of a nonleaf vertex has distribution which is dominated by the $B(n, d/n)$ with the condition that there are at least two children. Let $Z$ be a random variable distributed as in $B(n, d/n)$, clearly

$$Pr[Z \geq 2] = 1 - \left(1 - \frac{d}{n}\right)^n - n\frac{d}{n}\left(1 - \frac{d}{n}\right)^{n-1} \geq 1 - (d+1)e^{-d}$$

Let

$$a(i) = \begin{cases} \dfrac{t \cdot \mathcal{S}}{(\mathcal{S}-t)^2} & \text{if } i \leq t \\ \dfrac{\mathcal{S}}{q_{i,\mathcal{S},2}} & \text{otherwise} \end{cases}$$

where $q_{i,\mathcal{S},2}$, is as defined in (10). Consider the coupling $\mathcal{C}'(T, \mathcal{S}, l)$, where $T$ is an instance of $T_{r,d,\epsilon \log n}$ rooted at the vertex $r$. Let $E[p_r]$ be the expectation of the disagreement probability on the root $r$. Conditioning on the number of children of $r$ and the disagreement probability $p_w$, $\forall w \in CH_r$ in $\mathcal{C}'(T, \mathcal{S}, l)$, by Lemma B and Lemma C we have

$$E[p_r | p_w, \forall w \in CH_r] = a(|CH_r|) \sum_{w \in CH_r} p_w$$

By definition, $\forall w \in CH_r$, $p_w$ is upper bounded by the disagreement probability on the vertex $w$ in the coupling $\mathcal{C}'(T_w, \mathcal{S}, l-1)$ where $T_w$ is the subtree of $T$ rooted at vertex $w$. Call this disagreement probability $p_w^*$. We clear out that $p_w$ refers to the coupling $\mathcal{C}'(T, \mathcal{S}, l)$ while $p_w^*$ to $\mathcal{C}'(T_w, \mathcal{S}, l-1)$. It is direct that

$$E[p_r] \leq \sum_{i=0}^{n} i a(i) Pr[|CH_r| = i] E[p_w^*] \qquad \text{for } w \in CH_r \qquad (11)$$

Also, noting that the function $f(i) = i \cdot a(i)$ is increasing for $t << S$ and by the fact that the distribution of the number of children of $r$ is dominated by the $B(n, d/n)$, (by proposition 9.1.2. of [13]), it holds that

$$E[p_r] \leq \frac{1}{1-(d+1)e^{-d}} \sum_{i=0}^{n} a(i) \binom{n}{i} p^i (1-p)^{n-i} E[p_w^*] \qquad \text{for } w \in CH_r$$

where $p = d/n$.

Let $S_1 = \sum_{i=0}^{t} i \cdot a(i) \binom{n}{i} p^i (1-p)^{n-i}$ and $S_2 = \sum_{i=t+1}^{n} i \cdot a(i) \binom{n}{i} p^i (1-p)^{n-i}$. Using the derivation of Lemma D we have that

$$S_1 = \frac{t \cdot S}{(S-t)^2} q(t) d$$

Before calculating $S_2$, we eliminate the probability term $q_{i,S,2}$ from the $a(i)$ for $i > t$. Note that $q_{i,S,2} > q_{i-1,S-1}$, i.e. $q_{i-1,S-1}$ is the probability for not choosing all the elements of a set of cardinality $S - 1$ after $i - 1$ trials when at each trial we choose u.a.r. a member of the set. For $q_{i,S,2}$ it holds that

$$q_{i,S,2} \geq q_{i-1,S-1} \geq (S-1)\left(1 - \frac{1}{S-1}\right)^{i-1}\left(1 - q_{i-1,(S-2)}\right)$$

i.e. the probability of the event "not choosing *some* of the $S - 1$ elements after $i - 1$ trials" is greater than, or equal to the probability of the event "not choosing *exactly* one of $S - 1$ elements after $i - 1$ trials", since the second event is a special case of the first one. Furthermore, since $q_{i-1,(S-2)} \leq q_{i-1,S-1}$ we get that

$$q_{i-1,S-1} \geq (S-1)\left(1 - \frac{1}{S-1}\right)^{i-1}\left(1 - q_{i-1,S-1}\right) \tag{12}$$

Let $\Omega = \{1, \ldots, n\}$ and let $t_0 = \sup\{t \in \Omega | \ q_{t-1,S-1} \geq 1/2\}$. Instead of using $q_{i-1,S-1}$ we make the following simplification. For $i > t_0$ we bound the quantity $1/q_{i-1,S-1}$ as

$$\frac{1}{q_{i-1,S-1}} \leq \frac{1}{(S-1)\left(1 - \frac{1}{S-1}\right)^{i-1}} = \frac{2}{S-1}\left(\frac{S-1}{S-2}\right)^{i-1}$$

Also, for $i \leq t_0$, clearly, $1/q_{i,S} \leq 2$. With derivations similar to those in the proof of lemma D for $S_2$ we get that

$$S_2 \leq 2d \left( S(1 - q(t)) + \frac{S}{S-1} \left( \exp\{d/(S-2)\} - q(t) \right) \right)$$

Substituting the bounds for $S_1$ and $S_2$ in (11) we get

$$E[p_r] \leq \frac{1}{1-(d+1)e^{-d}} \left( d\frac{t \cdot S}{(S-t)^2} q(t) + 2d \left( S(1-q(t)) + \frac{S}{S-1} \left( \exp\{d/(S-2)\} - q(t) \right) \right) \right) E[p_w^*]$$

for $w \in CH_r$. We can substitute $E[p_w^*]$ in the same manner as $E[p_r]$. Using induction and assuming that for the vertices at distance $l$ from the root the expectation of the probability of disagreement is 1, the lemma follows. ◇

Lemma H, follows by combining the lemmas F and G.

**Lemma H** Consider a system $PCS(G_{v,d,\epsilon \log n}, \mathcal{S})$, for $d > 1$, $\epsilon = \frac{0.9}{4\log(e^2 d/2)}$ and for $G_{v,d,\epsilon \log n}$ we assume that it is a unicyclic graph. If the cardinality of $\mathcal{S}$ is a sufficiently large constant, then with probability at least $1 - 2n^{-1.25}$, for the above system it holds that $SD(v, \lfloor \epsilon \log n \rfloor) = n^{-1.25}$. For sufficiently large d, we should have $\mathcal{S} \geq d^{14}$.

**Proof:** To prove the lemma we first show that using the coupling $\mathcal{C}'(T_{r,d,\epsilon \log n}, \lfloor \epsilon \log n \rfloor)$ for the system $PCS(T_{r,d,\epsilon \log n}, \mathcal{S})$, it holds that $SD(r, \lfloor \epsilon \log n \rfloor) = n^{-1.25}$ with probability at least $1 - 2n^{-1.25}$, when $\mathcal{S}$ is a sufficiently large constant and for sufficiently large d, we should have $\mathcal{S} \geq d^{14}$. Then, the lemma will follow by using Lemma F.

By Lemma G we have that, for the coupling $\mathcal{C}'$ over the trees $T_{r,d,\epsilon \log n}$ when $\mathcal{S}$ colours are used and the boundary set $L$ is at distance, at least, $l$ from the root $r$, the expectation of $p_r$ is bounded as

$$E[p_r] \leq \left( \frac{1}{1-(d+1)e^{-d}} \left( d\frac{t \cdot \mathcal{S}}{(\mathcal{S}-t)^2} q(t) + 2d \left( \mathcal{S}(1-q(t)) + \frac{\mathcal{S}}{\mathcal{S}-1} (\exp\{d/(\mathcal{S}-2)\} - q(t)) \right) \right) \right)^l \quad (13)$$

where $l$ is the distance of $v$ and the boundary $L$, which in our case is $l = \lfloor \epsilon \log n \rfloor$, with $\epsilon = \frac{0.9}{\log(e^2 d/2)}$. Also, $q(t)$ is equal to the probability for a random variable, distributed as in $B(n-1, d/n)$, for fixed d, to be less than t, the maximum number of children of a mixing vertex.

Set $l = \epsilon \log n$, with $\epsilon = \frac{0.9}{\log(e^2 d/2)}$ in (8). To prove the lemma it suffices to show that for $\mathcal{S}$ as described in the statement of the lemma and appropriately large $t$ we have $E[p_r] \leq n^{-2.5}$. Clearly, for $E[p_r] \leq n^{-2.5}$ by using the Markov Inequality (see [3]) we can get that

$$Pr[p_r \geq 2n^{-1.25}] \leq \frac{E[p_r]}{2n^{-1.25}} = n^{-1.25}/2$$

By Definition F and Theorem D we get that if $E[p_r] \leq n^{-2.5}$, then with probability $1 - Pr[p_r \geq 2n^{-1.25}] \geq 1 - 2n^{-1.25}$ the system $PCS(T_{r,d,\epsilon \log n}, \mathcal{S})$ it holds $SD(r, \lfloor \epsilon \log n \rfloor) \leq n^{-1.25}$, which proves the lemma.

First we show that for sufficiently large d, for $\mathcal{S} \geq d^{14}$ and $t$ such that $q(t) \geq 1 - d^{-28}$ we get $E[p_r] \leq n^{-2.5}$. By Corollary B (in the proof of lemma E) we have that if a random variable $Z$ is distributed as in $B(n, q)$ with $\lambda = nq$ then

$$Pr[Z \geq x] \leq e^{-x} \quad x \geq 7\lambda. \quad (14)$$

Thus, for $t = \max\{7d, 28\log d + 1\}$ it holds $q(t) \geq 1 - d^{-28}$.

Assuming that d is a sufficiently large constant, we substitute $\mathcal{S}$ and $t$ in (13) and we get

$$E[p_r] \leq \left( \frac{1}{1-(d+1)e^{-d}} \left( \frac{7d^{16}}{(d^{14}-7d)^2} + \frac{2d}{1-d^{-14}} \left( d^{-14} + 1 + \frac{d}{d^{14}-1} + \frac{e^\xi}{2!}\frac{d^2}{(d^{14}-1)^2} - 1 + d^{-28} \right) \right) \right)^{\epsilon \log n}$$

where $0 < \xi < d/(d^{14}-1)$. In the above inequality we used the fact that $1 - d^{-28} \leq q(t) \leq 1$, and we substituted $e^{d/(\mathcal{S}-1)}$ by using the MacLaurin series of the function $f(x) = e^x$, for real $x$. Thus, we get

$$E[p_r] \leq \left( \frac{d^{-12}}{1-(d+1)e^{-d}} \left( \frac{7}{(1-7d^{-13})^2} + \frac{2d^{-1}}{1-d^{-14}} + \frac{2}{(1-d^{-14})^2} + \frac{ed^{-13}}{(1-d^{-14})^3}\frac{+2d^{-15}}{1-d^{-14}} \right) \right)^{\epsilon \log n}$$

Taking $d$ at least 20, we get that

$$E[p_r] \leq n^{\epsilon \log(9.2d^{-12})}$$

Replacing $\epsilon$, we see that it suffices to hold $0.9 \log(9.2d^{-12}) \leq -10 \log(e^2 d/2)$, or $9.2d^{-12} \leq (e^2 d/2)^{-11.11}$ which clearly holds for sufficiently large constant $d$.

For relatively smaller $d$, one can easily see that setting $\mathcal{S} = d^x$, for an appropriately large constant exponent $x$ and arranging the quantity $t$ so as $q(t) \geq 1 - d^{-2x}$ similarly to the previous case, can get

$$E[p_r] \leq \left( \frac{d^{-x+2}}{1-(d+1)e^{-d}} \left( \frac{t/d}{(1-d^{-x}t)^2} + \frac{2d^{-1}}{1-d^{-x}} + \frac{2d^{-x-1}}{1-d^{-x}} + \frac{2}{(1-d^{-x})^2} + e\frac{d^{-x+2}}{(1-d^{-x})^3} \right) \right)^{\epsilon \log n}$$

We take $x$ sufficiently large so as to have $1 - d^{-x} \geq 1 - 10^{-3}$ and $xd^{-x} \leq 10^{-3}$.

If $t = 7d$, then, with the above assumptions, about $x$, we can easily derive that $E[p_r] \leq (d^{-x+2} 44)^{\epsilon \log n}$. For this case to have $E[p_r] \leq n^{-2.5}$ we should have $44d^{-x+2} \leq (e^2 d/2)^{11.11}$, which clearly holds for sufficiently large $x$.

If $2x \log d + 1 > 7d$, then by Corollary B we should have $t = 2x \log d + 1$. With the assumptions we have made for $x$ we get that $E[p_v] \leq (d^{-x+2}(8x\frac{\log d}{d} + 16))^{\epsilon \log n}$. Thus, if $E[p_r] \leq n^{-2.5}$, then we should have $(d^{-x+2}(8x\frac{\log d}{d} + 16)) \leq (e^2 d/2)^{11.11}$, which clearly holds for sufficiently large $x$.

The lemma follows by using Lemma F.

◇

## 4 Properties of the algorithm

We close this work by restating and proving the two theorems that deal with the issues of accuracy and efficiency of the sampling algorithm.

**Theorem B** *If $\mathcal{S}$ is a sufficiently large integer constant, then, with probability $1 - O(n^{-0.1})$, the sampling algorithm is successful and returns a $\mathcal{S}$-colouring of the input graph $G$, whose distribution is within total variation distance $n^{-0.25}$ from the uniform over all the proper $\mathcal{S}$-colourings of $G$.*

**Proof:** The algorithm is considered successful if the spin-system it considers has the properties stated in section 2.1, i.e. the following two hold: *First*, for each iteration $i$, the induced subgraph, of the input graph, that contains $v_i$ and all the vertices that are within graph distance $\lfloor \epsilon \log n \rfloor$ from $v_i$, with $\epsilon = \frac{0.9}{4 \log(de^2/2)}$, is either unicyclic or a tree. According to Lemma A this holds with probability at least $1 - n^{-0.1}$. *Second*, the spatial mixing property stated in Theorem A holds. For sufficiently large constant $\mathcal{S}$, the spatial mixing property stated in Theorem A holds with probability at least $1 - n^{-0.25}$.

Consider that $G$ is the input graph of the algorithm, which is an instance of $G_{n,d/n}$ and we take $\mathcal{S}$ as large as indicated in Theorem A. The, the algorithm is successful with probability at least $1 - (n^{-0.1} + 2n^{-0.25}) = 1 - O(n^{-0.1})$. From now on, we assume that the input $G$, belongs to this set of instances of $G_{n,d/n}$ that the algorithm is successful (which includes almost all instances, for sufficiently large $\mathcal{S}$).

What remains to be shown is the bound for the total variation distance between the distributions of the colouring that is returned by the algorithm and the uniform over all the proper $\mathcal{S}$-colouring of the input graph $G$.

First we show that (1) is valid. Consider two systems, i.e. $S_1 = PCS(G, \mathcal{S})$ and $S_2 = PCS(G, \mathcal{S})$, each independent of the other. Assume that, in both systems, we fix the colour assignments of the vertices in $A_i$, according to probability measure $\mu(\cdot)$. Clearly, we have for both systems that $Pr[X_{A_i} = C(A_i)] = \mu(X_{A_i} = C(A_i))$, for $C(A_i) \in [\mathcal{S}]^{A_i}$. After fixing the colour assignments of the vertices in $A_i$ we look at the colour assignments of the vertices at graph distance, at least, $\lfloor \epsilon \log n \rfloor$ from the vertex $v_i$, in both systems. Let $V'$ be the vertex set whose colouring is revealed and let $C(V')$ be the colouring we see in $S_1$ and let $C'(V')$ be the colouring we see in $S_2$.

By the law of total probability, it is easy, for one, to see that, in $S_1$, the probability for $v_i$ to be assigned colouring $c$, $\forall c \in [\mathcal{S}]$, is equal to $\mu(X_{v_i} = c | C_1(V'))$. Similarly in $S_2$, the the probability for $v_i$ to be assigned colouring $c$, $\forall c \in [\mathcal{S}]$, is given by the measure $\mu(X_{v_i} = c | C_1(V'))$.

Note that in the above we have fixed the colourings of the vertices according to $\mu(\cdot)$, but we do not have seen what are the actual colourings. We see, only, the colourings of the vertices in $V'$.

By Theorem A and the discussion at the end of the section 1.2, we have that the total variation distance between the probability measures of interest, is bounded as follows:

$$d_{TV}(\mu(X_{v_i}|C_1(V')), \mu(X_{v_i}|C_2(V'))) \leq SD(v_i, \lfloor \epsilon \log n \rfloor) \leq n^{-1.25} \quad (15)$$

Which shows that (1) is valid. Thus, the *asymptotic independence* between the colouring of the vertex $v_i$ and the colouring of the vertex set $V'$ remains when we colour the vertices in $A_i$ according to distribution $\mu(\cdot)$.

Consider the following coupling of our algorithm and an ideal algorithm that gives a perfect uniform sample by colouring vertices one by one in some way. At each repetition, both algorithms assign a colour to some (the same) vertex in $G_{n,d/n}$. Consider a specific step of the coupling where the vertex $v$ is to be coloured. By Theorem A and (1), we can have a sufficiently large $\mathcal{S}$ such that, conditioning on the fact that all vertices until now are identically coloured by the two algorithms, the probability for $v$ to have a different colour assignment in the coupling is at most $n^{-1.25}$. Thus, the probability for the coupling to end with no disagreement is at least $(1 - n^{-1.25})^n > 1 - n^{-0.25}$. The theorem follows. ◇

**Theorem C** *The time complexity of the sampling algorithm is w.h.p. asymptotically upper bounded by $O(n^2)$, where $n$ is the number of vertices of the input graph.*

**Proof:** First, we note that the algorithm will return failure if any of the graphs $B_{v_i}$ is neither unicyclic, nor a tree. The number of step the algorithm needs, in the case of failure, is at most equal to the number of steps that are need in the case of nonfail. Thus, time complexity of the nonfailing execution is an upper bound of the time complexity of the algorithm.

The algorithm needs $O(n)$ steps to create the graph $B_{v_i}$, at the $i$-th iteration of its for loop. The graph $B_{v_i}$ can be created by using any traversal algorithm, e.g. depth first search. This time bound follows by the fact that the number of vertices and the number of vertices in $B_{v_i}$ are upper bounded by the number of vertices and the edges of the input graph. Using the Chernoff bounds (see [6]) it is direct to show that w.h.p. the number of edges in an instance of $G_{n,d/n}$ is $O(n)$.

Implementing a colouring of $v_i$ according to $\mu(X_{v_i}|C(A_i \cap V_i))$, is equivalent to generating a random list colouring of $B_{v_i}$ and keeping only the colour assignment of $v_i$ from this colouring. In the list colouring problem every vertex $u \in V_i$ has a set $L(u)$ of valid colours, where $L(u) \subseteq [\mathcal{S}]$ and $u$ can only receive a colour in $L(u)$. As argued in [4], for a tree on $l$ vertices we can exactly compute the number of $k$-colourings in time $l \cdot k$. Therefore we can also generate a random list colouring of the tree. Also, for a unicyclic component we can simply consider all the $\leq k^2$ colourings of the endpoints of the extra edge and then recurse the remaining tree.

I.e. the time we need to colour $v_i$ according to $\mu(X_{v_i}|C(A_i \cap V_i))$, is at most $O(n)$. The theorem follows by noting that we need to colour $n$ vertices.

◇

**Acknowledgement.** We wish to thank D. Achlioptas for comments on our work and also E. Mossel and A. Sly for the exchange of comments about our and their proving methodologies. We wish to thank previous anonymous referees for comments that allowed us to structure the proof arguments better.